\documentclass[prd,nofootinbib,showpacs,preprint]{revtex4}
\usepackage{amsmath}
\usepackage{graphicx}
\usepackage{epsfig}
\graphicspath{{Figs/}}
\usepackage{dcolumn}
\usepackage{bm}
\usepackage{amssymb}
\usepackage[usenames,dvipsnames]{color}
\usepackage{slashed}
\usepackage[dvipdfm,colorlinks,citecolor=blue]{hyperref}
\usepackage{booktabs}
\usepackage{multirow}
\begin{document}
\title{Connecting Leptonic CP Violation, Lightest Neutrino Mass and Baryon Asymmetry Through Type II Seesaw}
\author{Rupam Kalita}
\email{rup@tezu.ernet.in}
\author{Debasish Borah}
\email{dborah@tezu.ernet.in}
\affiliation{Department of Physics, Tezpur University, Tezpur-784028, India}

\begin{abstract}
We study the possibility of connecting leptonic Dirac CP phase $\delta$, lightest neutrino mass and baryon asymmetry of the Universe within the framework of a model where both type I and type II seesaw mechanisms contribute to neutrino mass. Type I seesaw gives rise to Tri-Bimaximal (TBM) type neutrino mixing whereas type II seesaw acts as a correction in order to generate non-zero $\theta_{13}$. We derive the most general form of type II seesaw mass matrix which can not only give rise to correct neutrino mixing angles but also can generate non-trivial value of $\delta$. Considering both the cases where type II seesaw is sub-leading and is equally dominant compared to type I seesaw, we correlate the type II seesaw term with $\delta$ and lightest neutrino mass. We further constrain the Dirac CP phase $\delta$ and hence the type II seesaw mass matrix from the requirement of producing the observed baryon asymmetry through the mechanism of leptogenesis.

\end{abstract}
\pacs{12.60.-i,12.60.Cn,14.60.Pq}
\maketitle
\section{Introduction}
\label{intro}
Neutrino oscillation experiments \cite{PDG,PDG2,PDG3,PDG4,PDG5} in the last few years have given substantial amount of evidence in favor of non-zero neutrino mass and mixing. More recently, neutrino oscillation experiments T2K \cite{T2K}, Double ChooZ \cite{chooz}, Daya-Bay \cite{daya} and RENO \cite{reno} 
have not only made the earlier predictions for neutrino parameters more precise, but also predicted non-zero value of the 
reactor mixing angle $\theta_{13}$. The latest global fit value for $3\sigma$ range of neutrino oscillation parameters \cite{schwetz12} are as follows:
$$ \Delta m_{21}^2=(7.00-8.09) \times 10^{-5} \; \text{eV}^2$$
$$ \Delta m_{31}^2 \;(\text{NH}) =(2.27-2.69)\times 10^{-3} \; \text{eV}^2 $$
$$ \Delta m_{23}^2 \;(\text{IH}) =(2.24-2.65)\times 10^{-3} \; \text{eV}^2 $$
$$ \text{sin}^2\theta_{12}=0.27-0.34 $$
$$ \text{sin}^2\theta_{23}=0.34-0.67 $$ 
\begin{equation}
\text{sin}^2\theta_{13}=0.016-0.030
\label{equation:data1}
\end{equation}
Another global fit study \cite{fogli} reports the 3$\sigma$ values as
$$ \Delta m_{21}^2=(6.99-8.18) \times 10^{-5} \; \text{eV}^2$$
$$ \Delta m_{31}^2 \;(\text{NH}) =(2.19-2.62)\times 10^{-3} \; \text{eV}^2 $$
$$ \Delta m_{23}^2 \;(\text{IH}) =(2.17-2.61)\times 10^{-3} \; \text{eV}^2 $$
$$ \text{sin}^2\theta_{12}=0.259-0.359 $$
$$ \text{sin}^2\theta_{23}=0.331-0.637 $$ 
\begin{equation}
\text{sin}^2\theta_{13}=0.017-0.031
\label{equation:data2}
\end{equation}
where NH and IH refers to normal and inverted hierarchies respectively. Although the $3\sigma$ range for the Dirac CP phase is $0-2\pi$, there are two possible best fit values of it found in the literature: $5\pi/3$ \cite{schwetz12} and $\pi$ \cite{fogli}.

Before the discovery of non-zero $\theta_{13}$, the neutrino oscillation data were consistent with the so called TBM form of the neutrino mixing matrix which predicts the mixing angles as $\theta_{12} \simeq 35.3^o, \; \theta_{23} = 45^o$ and $\theta_{13} = 0$. Such a mixing pattern studied extensively in the literature \cite{Harrison,Harrison2,Harrison3,Harrison4,Harrison5,Harrison6} however, needs to be corrected in order to accommodate non-zero $\theta_{13}$ discovered recently. Several interesting proposals \cite{nzt13,nzt132,nzt133,nzt134,nzt135,nzt136,nzt137,nzt138,nzt13GA} have already been put forward which include different corrections to TBM mixing in order to generate non-zero $\theta_{13}$. In the present work, we attempt to find a mechanism which can not only generate non-zero $\theta_{13}$ but can also shed some light on those parameters of the neutrino sector which are not yet accurately measured namely, the lightest neutrino mass and the leptonic Dirac CP phase. Apart from incorporating the constraints from neutrino oscillation experiments, we also take the constraints from cosmology into account. Cosmology can constrain the the sum of absolute neutrino masses as $\sum_i \lvert m_i \rvert < 0.23$ eV \cite{Planck13}. We can further constrain the neutrino parameters from cosmology if we assume leptonic sector origin of matter antimatter asymmetry of the Universe. The matter antimatter asymmetry or the baryon asymmetry of the Universe is measured in terms of baryon to photon ratio which according to the latest data available from Planck mission \cite{Planck13} is given as 
\begin{equation}
Y_B \simeq (6.065 \pm 0.090) \times 10^{-10}
\label{barasym}
\end{equation} 
This baryon asymmetry can be naturally generated through the well known mechanism of leptogenesis. According to this mechanism, the observed baryon asymmetry of the Universe is generated by generating an asymmetry in the leptonic sector first and later converting it into baryon asymmetry through electroweak sphaleron transitions \cite{sphaleron}. As pointed out first by Fukugita and Yanagida \cite{fukuyana}, the out of equilibrium CP violating decay of heavy Majorana neutrinos provides a natural way to create the required lepton asymmetry. A very good review of this mechanism can be found in \cite{davidsonPR}. 

In some of our recent works \cite{db-t2,db-t22,dbijmpa,dbmkdsp}, we carried out an exercise of generating non-zero $\theta_{13}$ by considering a model where both type I \cite{ti,ti2,ti3,ti4,ti5} and type II \cite{tii,tii2,tii3,tii4,tii5,tii6,tii7} seesaw are present. The type I seesaw was assumed to give TBM type neutrino mixing whereas type II seesaw acts as a perturbation in order to generate non-zero value of reactor mixing angle. Similar attempts to study the deviations from TBM mixing by using the interplay of two different seesaw mechanisms were also done in \cite{devtbmt2,devtbmt22,devtbmt23,devtbmt24,devtbmt25}. In our earlier works, we considered type I seesaw mass matrix as TBM type and assumed a minimal structure of the type II seesaw mass matrix required to break the $\mu-\tau$ symmetry associated with the TBM mixing. We also assumed the type I seesaw mass matrix to be of leading order by fitting it with the best fit values of neutrino mass squared differences and TBM mixing angles. One difficulty that we face within such framework is to control or restrict the form and strength of type II seesaw mass matrix in such a way that while generating the correct value of $\theta_{13}$, the mass squared differences as well as other mixing angles remain within the allowed range. Here we generalize our earlier works by dropping these two assumptions: the minimal $\mu-\tau$ symmetry breaking form of the type II seesaw mass matrix and sub-leading approximation for the type II seesaw term. We rather derive the most general type II seesaw mass matrix which can give rise to the correct value of non-zero $\theta_{13}$ and could also contain the leptonic Dirac CP phase. We also consider three different scenarios where type I seesaw contribution to neutrino mass can be $50\%, 70\%$ or $90\%$ which includes the case where type I and type II seesaw are equally contributing to the neutrino mass. We constrain the type II seesaw mass matrix from the requirement of generating successful neutrino oscillation data and also study the correlation among leptonic Dirac CP phase, lightest neutrino mass and type II seesaw strength. We further constrain the leptonic Dirac CP phase contained in the type II seesaw mass matrix from the requirement of producing the observed baryon asymmetry through leptogenesis. It should be noted that throughout our study, we have not included Majorana neutrino phases in our calculations and hence the conclusions we arrive at in this work are valid only when the Majorana phases take extremal values $0$ or $2\pi$.

This paper is organized as follows. In section \ref{sec:seesaw}, we briefly discuss type II seesaw. In section \ref{sec:devtbm}, we discuss the deviations from TBM mixing using type II seesaw. In section \ref{sec:numeric} we describe the numerical analysis adopted here and finally conclude in section \ref{sec:conclude}.

\section{Type II Seesaw}
\label{sec:seesaw}
Type II seesaw mechanism is the extension of the standard model with a scalar field $\Delta_L$ which transforms like a triplet under $SU(2)_L$ and has $U(1)_Y$ charge twice that of lepton doublets. Such a choice of gauge structure allows an additional Yukawa term in the Lagrangian given by $ f_{ij}\ \left(\ell_{iL}^T \ C \ i \sigma_2 \Delta_L \ell_{jL}\right)$. The triplet can be represented as 
\begin{equation}
\Delta_L =
\left(\begin{array}{cc}
\ \delta^+_L/\surd 2 & \delta^{++}_L \\
\ \delta^0_L & -\delta^+_L/\surd 2
\end{array}\right) \nonumber
\end{equation} 
The scalar Lagrangian of the standard model also gets modified after the inclusion of this triplet. Apart from the bilinear and quartic coupling terms of the triplet, there is one trilinear term as well involving the triplet and the standard model Higgs doublet. From the minimization of the scalar potential, the neutral component of the triplet is found to acquire a vacuum expectation value (vev) given by 
\begin{equation}
 \langle \delta^0_L \rangle = v_L = \frac{\mu_{\Delta H}\langle \phi^0 \rangle^2}{M^2_{\Delta}}
\label{vev} 
\end{equation}
where $\phi^0=v$ is the neutral component of the electroweak Higgs doublet with vev approximately $10^2$ GeV. The trilinear coupling term $\mu_{\Delta H}$ and the mass term of the triplet $M_{\Delta}$ can be taken to be of same order. Thus, $M_{\Delta}$ has to be as high as $10^{14}$ GeV to give rise to tiny neutrino masses without any fine-tuning of the dimensionless couplings $f_{ij}$.

\begin{figure}[ht]
 \centering
\includegraphics[width=1.0\textwidth]{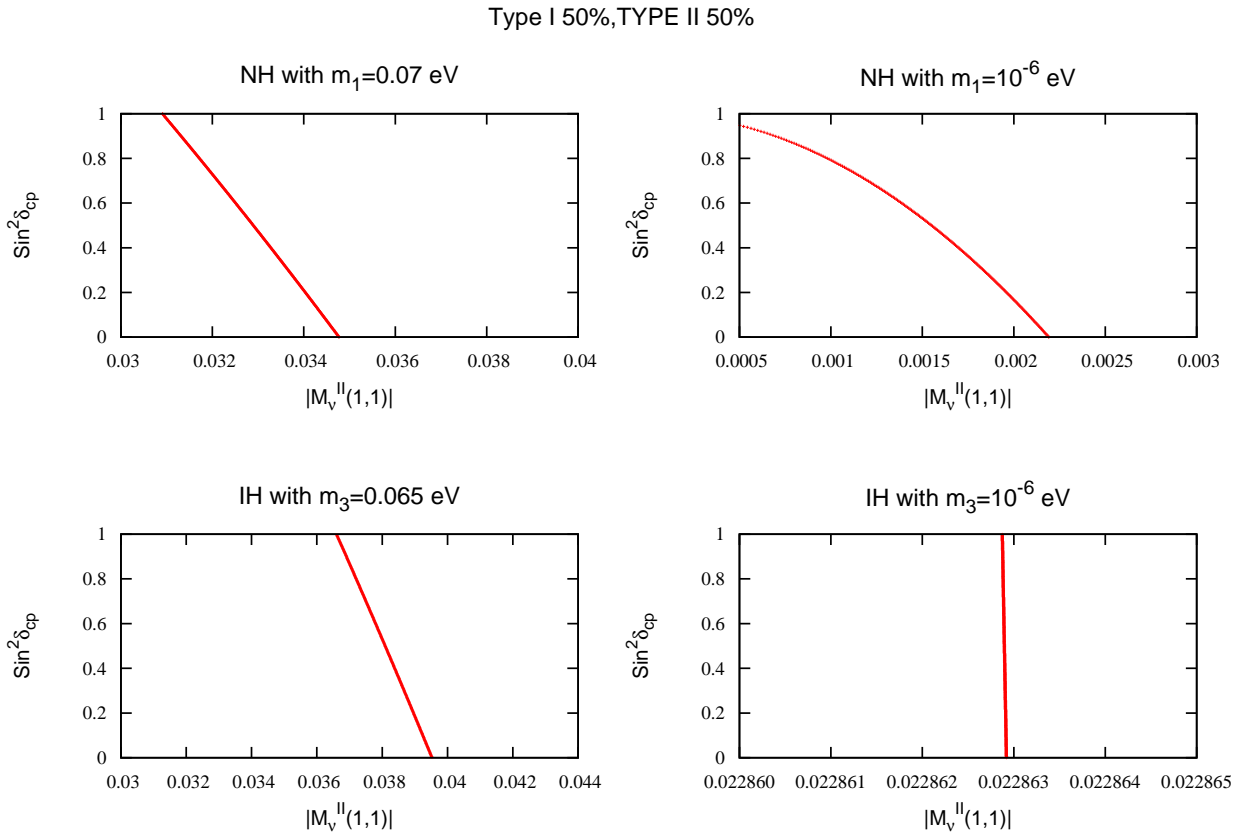}
\caption{Variation of $\sin^2\delta_{CP}$ with type II seesaw for $50\%$ contribution of type I seesaw to neutrino mass}
\label{fig1}
\end{figure}
\begin{figure}[ht]
 \centering
\includegraphics[width=1.0\textwidth]{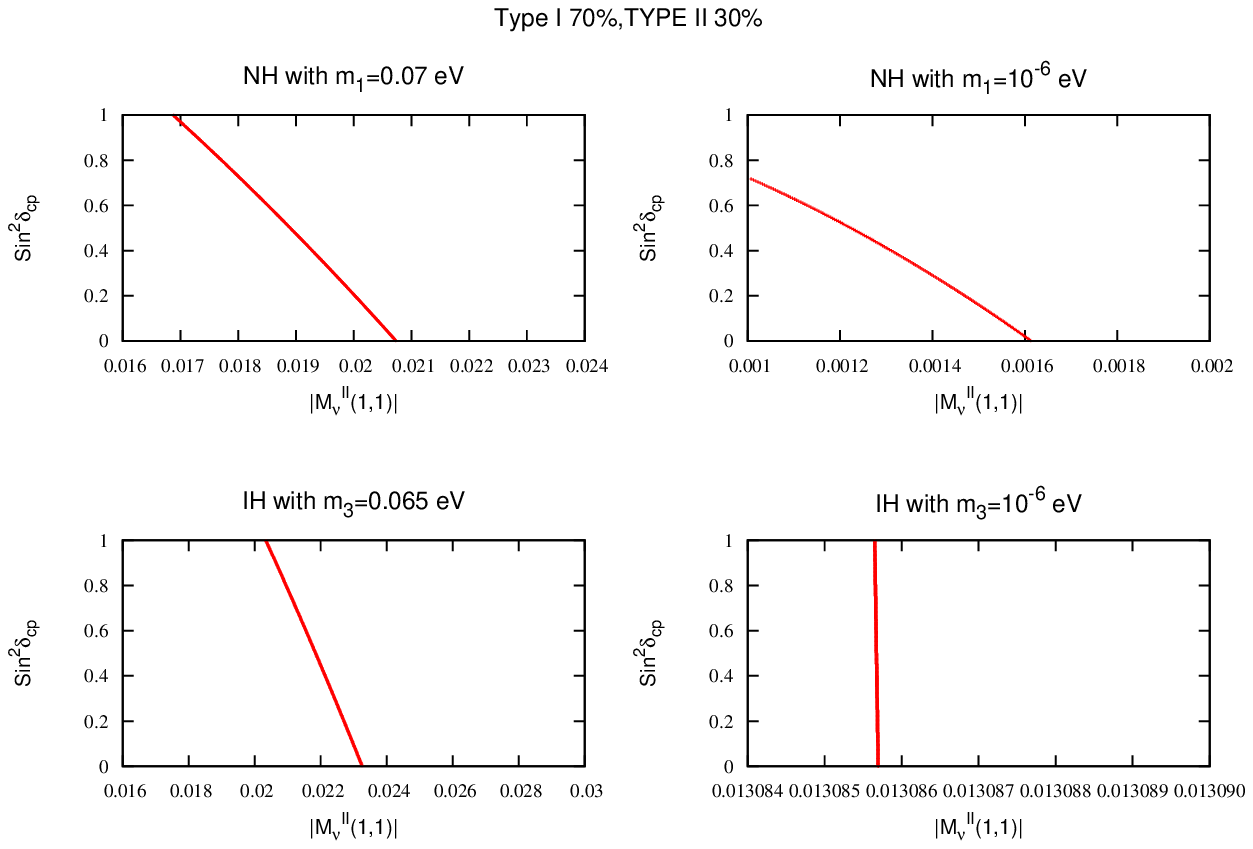}
\caption{Variation of $\sin^2\delta_{CP}$ with type II seesaw for $70\%$ contribution of type I seesaw to neutrino mass}
\label{fig2}
\end{figure}
\begin{figure}[ht]
 \centering
\includegraphics[width=1.0\textwidth]{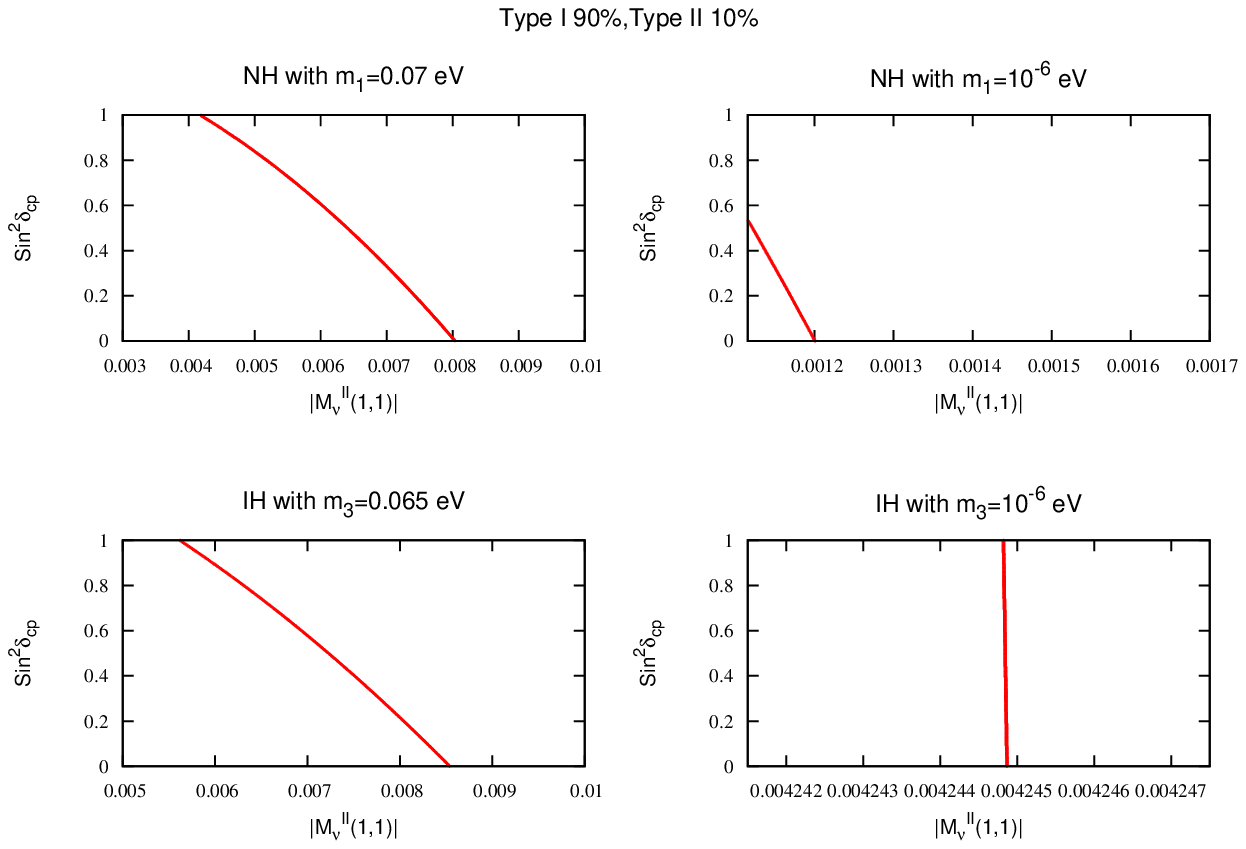}
\caption{Variation of $\sin^2\delta_{CP}$ with type II seesaw for $90\%$ contribution of type I seesaw to neutrino mass}
\label{fig3}
\end{figure}
\begin{figure}[ht]
\begin{center}
$
\begin{array}{cc}
\includegraphics[width=0.5\textwidth]{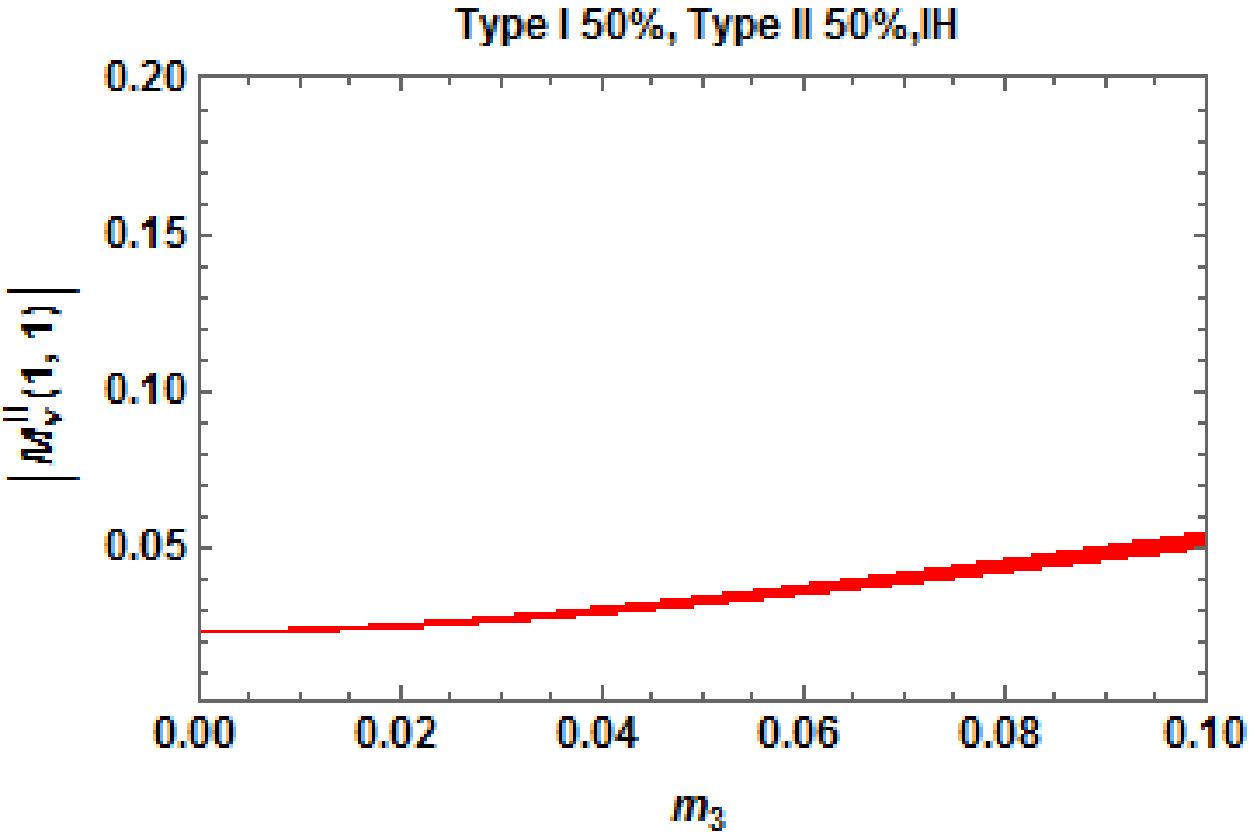} &
\includegraphics[width=0.5\textwidth]{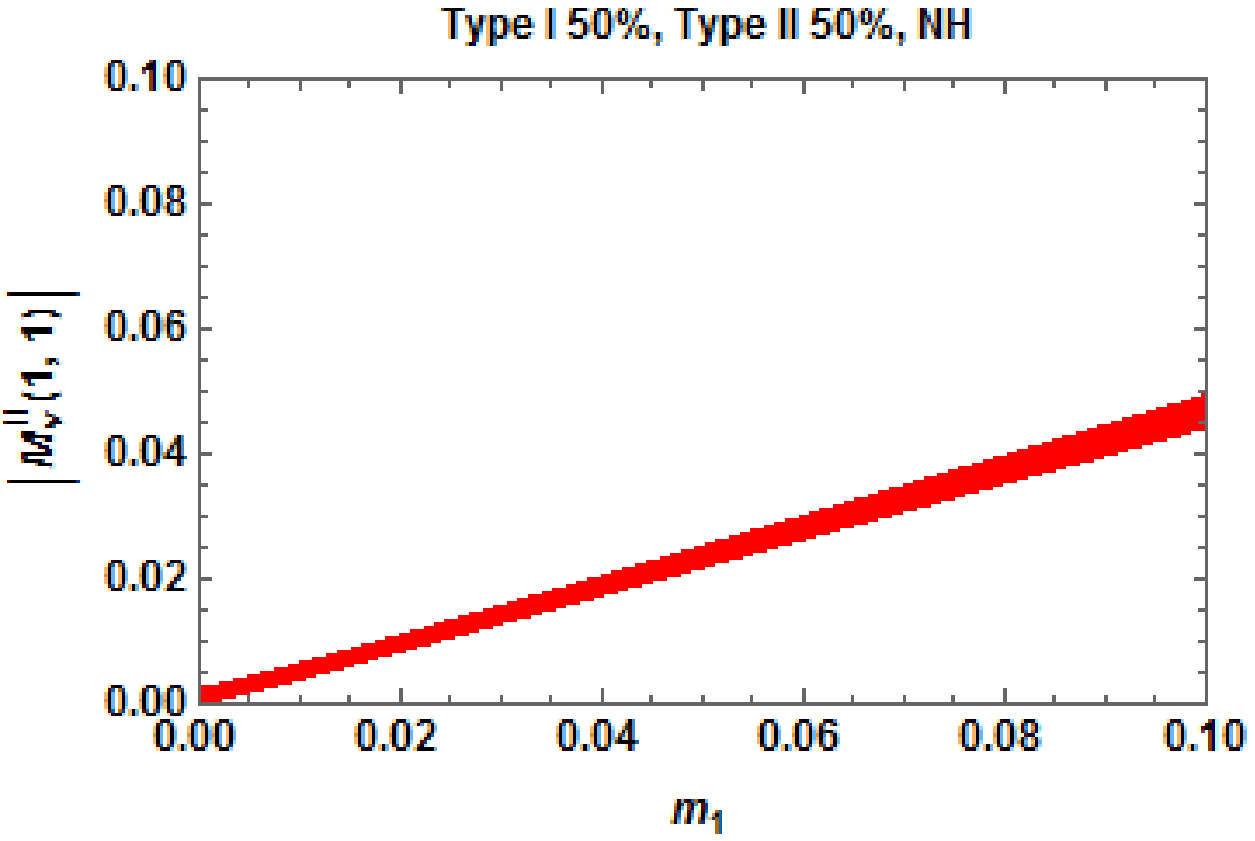}
\end{array}
$
\end{center}
\caption{Allowed parameter space in the type II seesaw mass term and lightest neutrino mass plane for $50\%$ contribution of type I seesaw to neutrino mass}
\label{fig4}
\end{figure}
\begin{figure}[ht]
\begin{center}
$
\begin{array}{cc}
\includegraphics[width=0.5\textwidth]{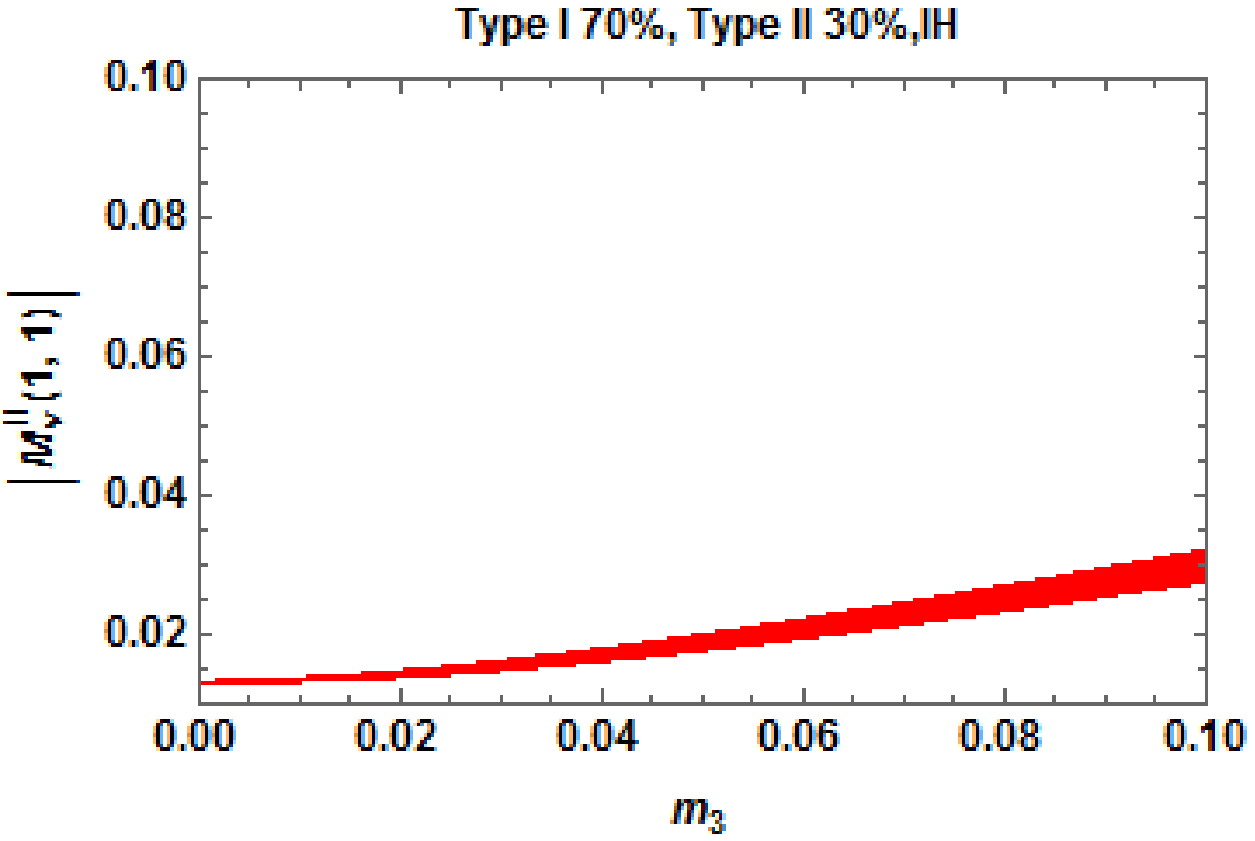} &
\includegraphics[width=0.5\textwidth]{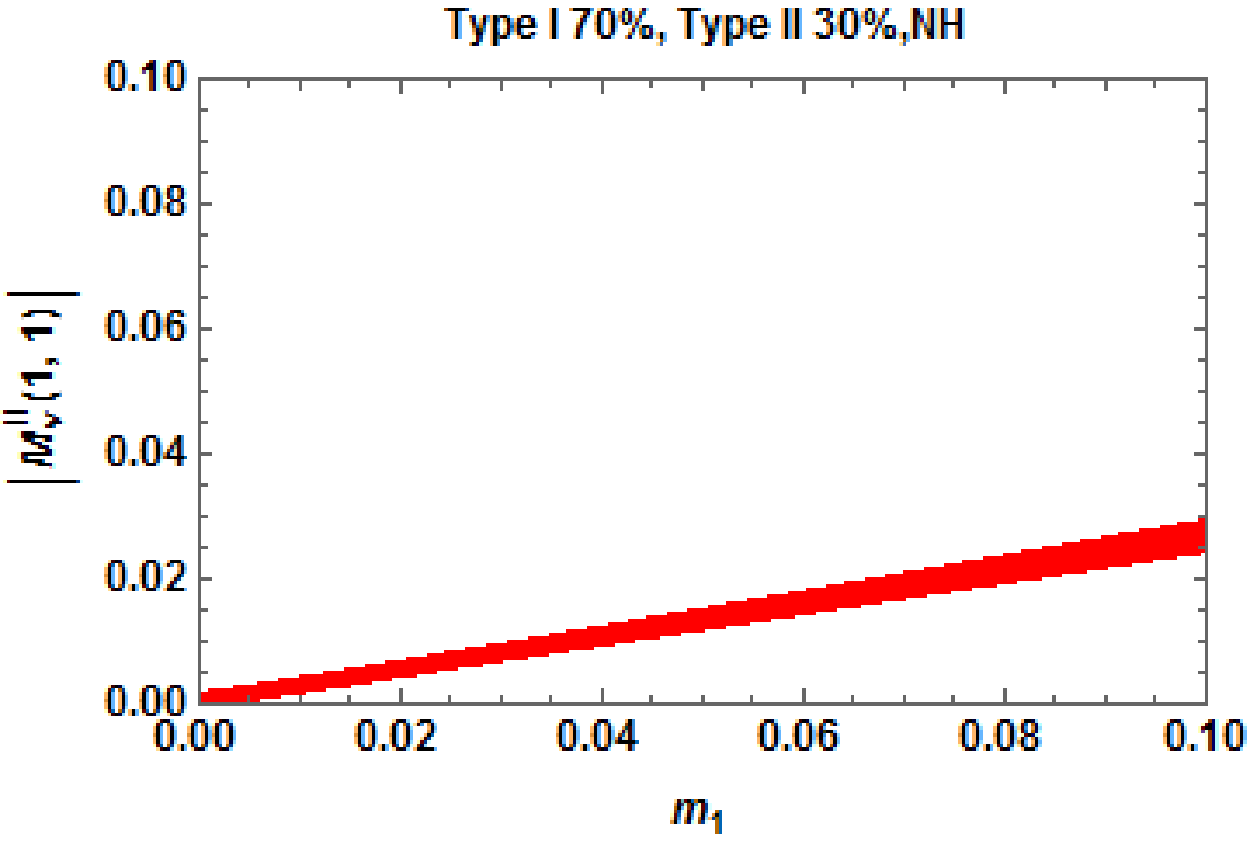}
\end{array}
$
\end{center}
\caption{Allowed parameter space in the type II seesaw mass term and lightest neutrino mass plane for $70\%$ contribution of type I seesaw to neutrino mass}
\label{fig5}
\end{figure}
\begin{figure}[ht]
\begin{center}
$
\begin{array}{cc}
\includegraphics[width=0.5\textwidth]{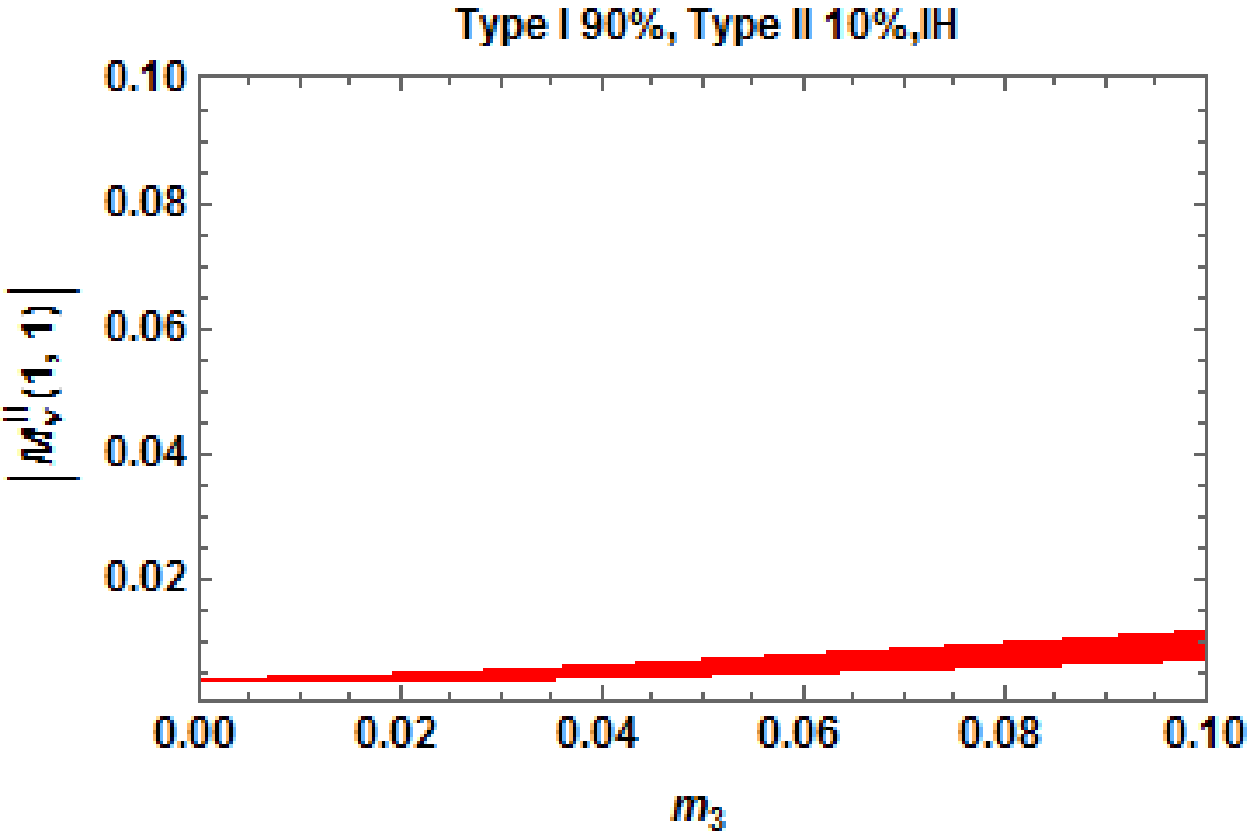} &
\includegraphics[width=0.5\textwidth]{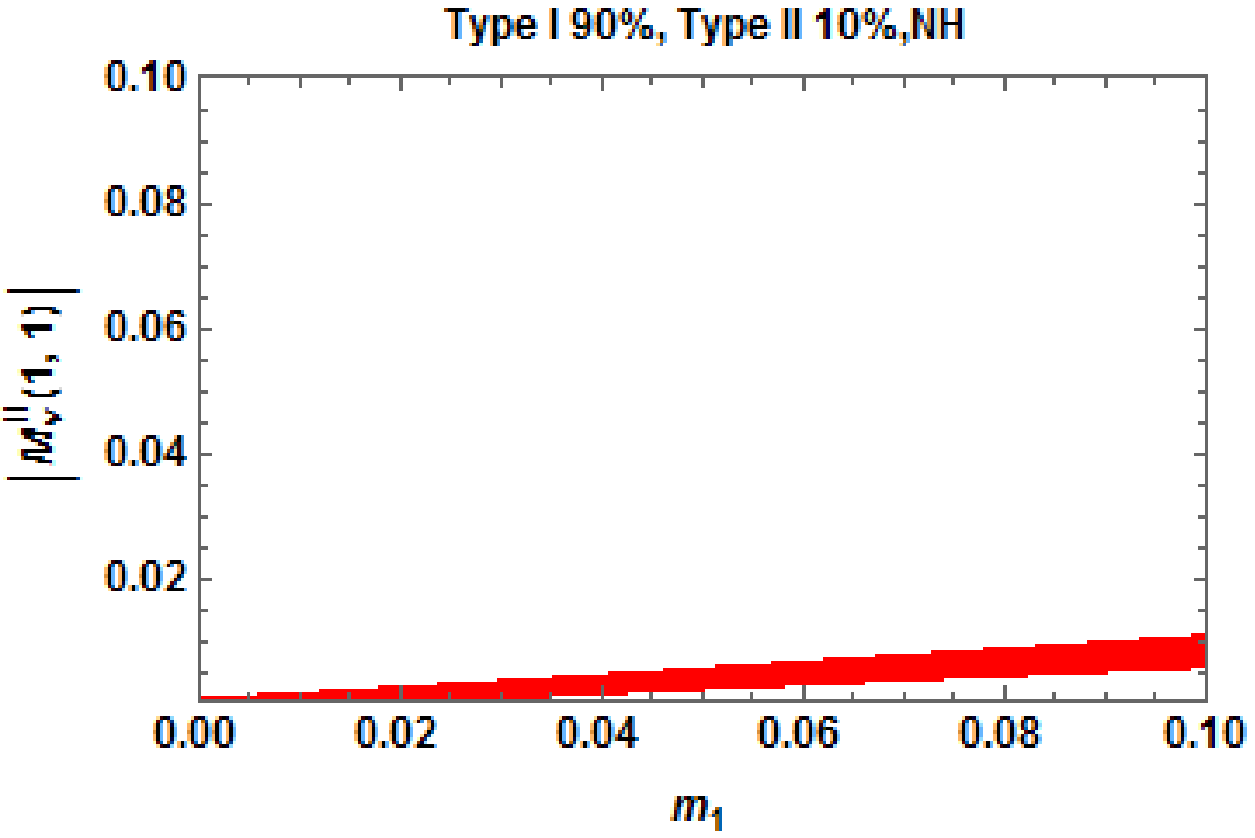}
\end{array}
$
\end{center}
\caption{Allowed parameter space in the type II seesaw mass term and lightest neutrino mass plane for $90\%$ contribution of type I seesaw to neutrino mass}
\label{fig6}
\end{figure}
\begin{figure}[ht]
 \centering
\includegraphics[width=0.85\textwidth]{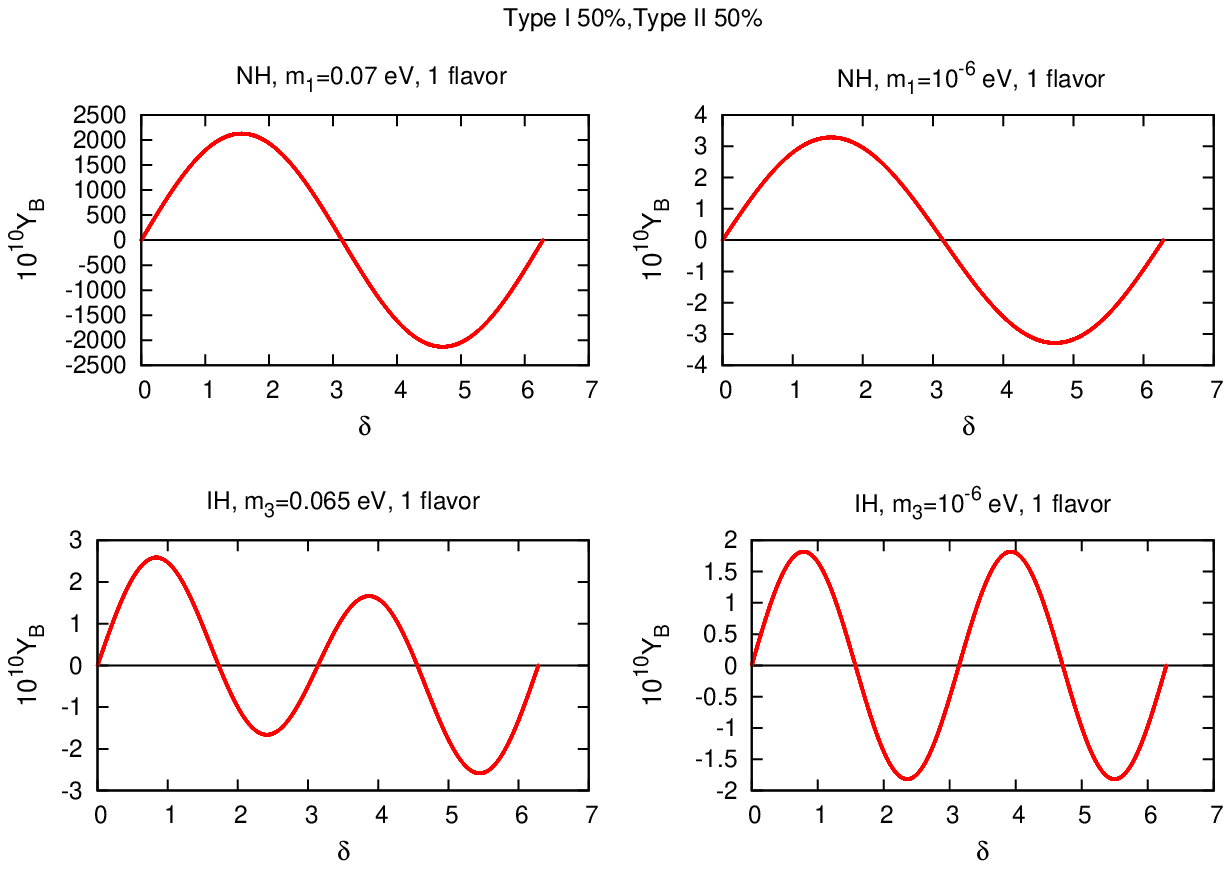}
\caption{Baryon asymmetry as a function of leptonic Dirac CP phase in the one flavor regime for $50\%$ contribution of type I seesaw to neutrino mass}
\label{fig7}
\end{figure}
\begin{figure}[ht]
 \centering
\includegraphics[width=0.85\textwidth]{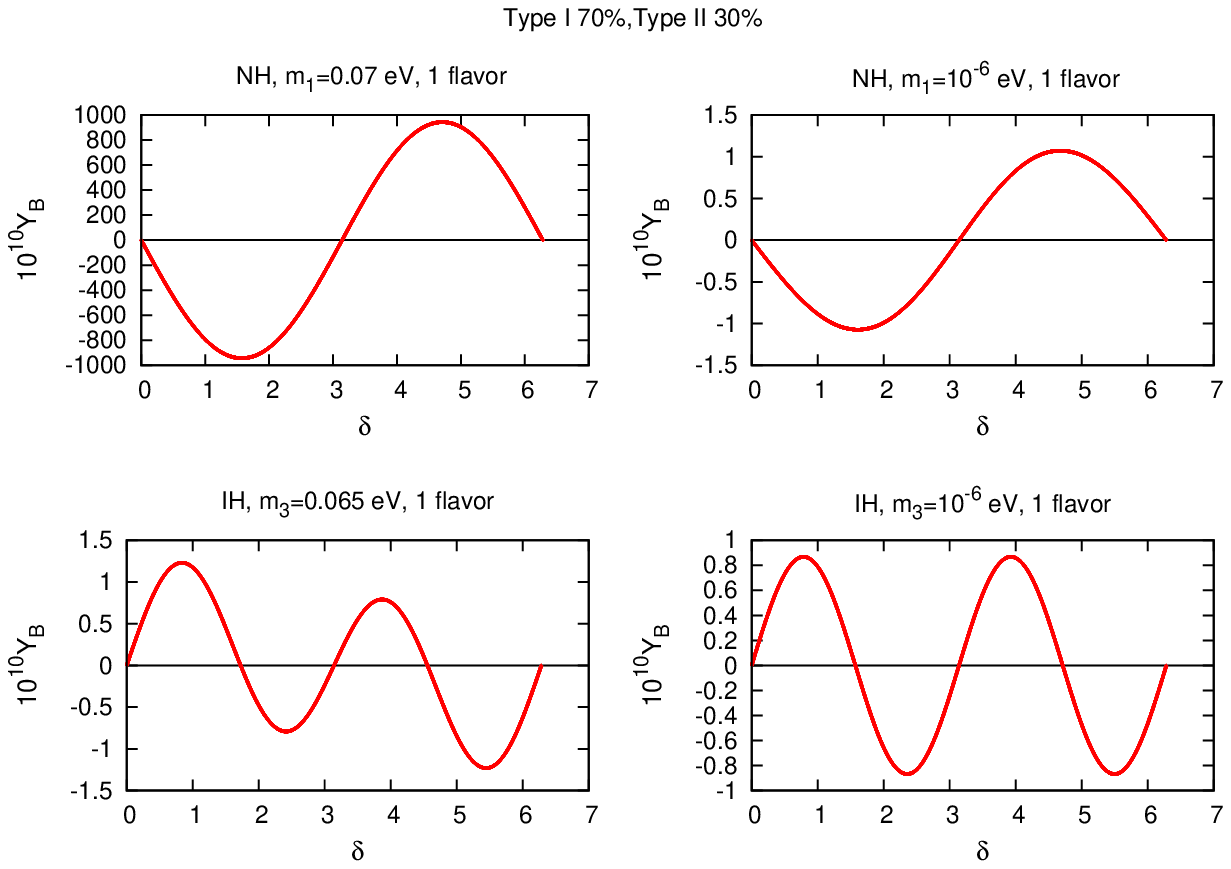}
\caption{Baryon asymmetry as a function of leptonic Dirac CP phase in the one flavor regime for $70\%$ contribution of type I seesaw to neutrino mass}
\label{fig8}
\end{figure}
\begin{figure}[ht]
 \centering
\includegraphics[width=0.85\textwidth]{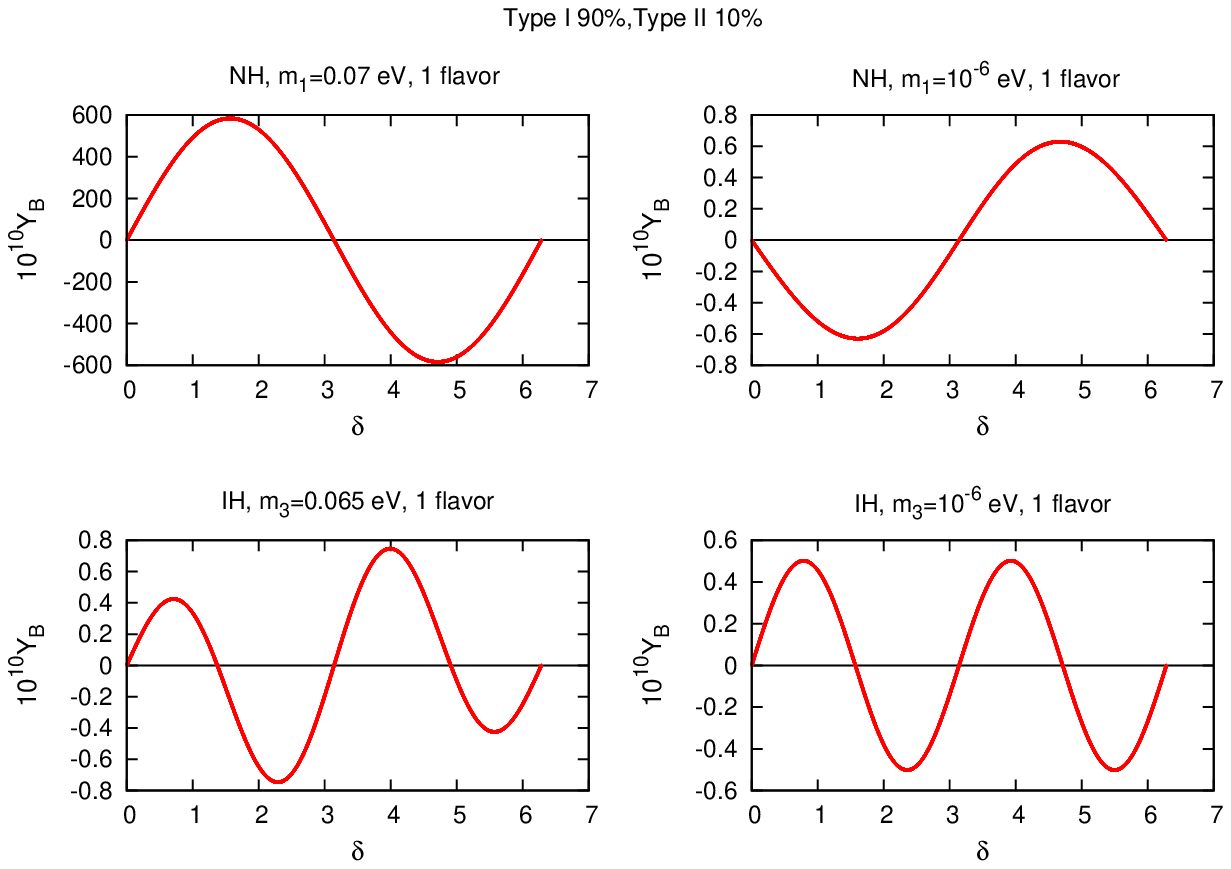}
\caption{Baryon asymmetry as a function of leptonic Dirac CP phase in the one flavor regime for $90\%$ contribution of type I seesaw to neutrino mass}
\label{fig9}
\end{figure}
\begin{figure}[ht]
 \centering
\includegraphics[width=0.85\textwidth]{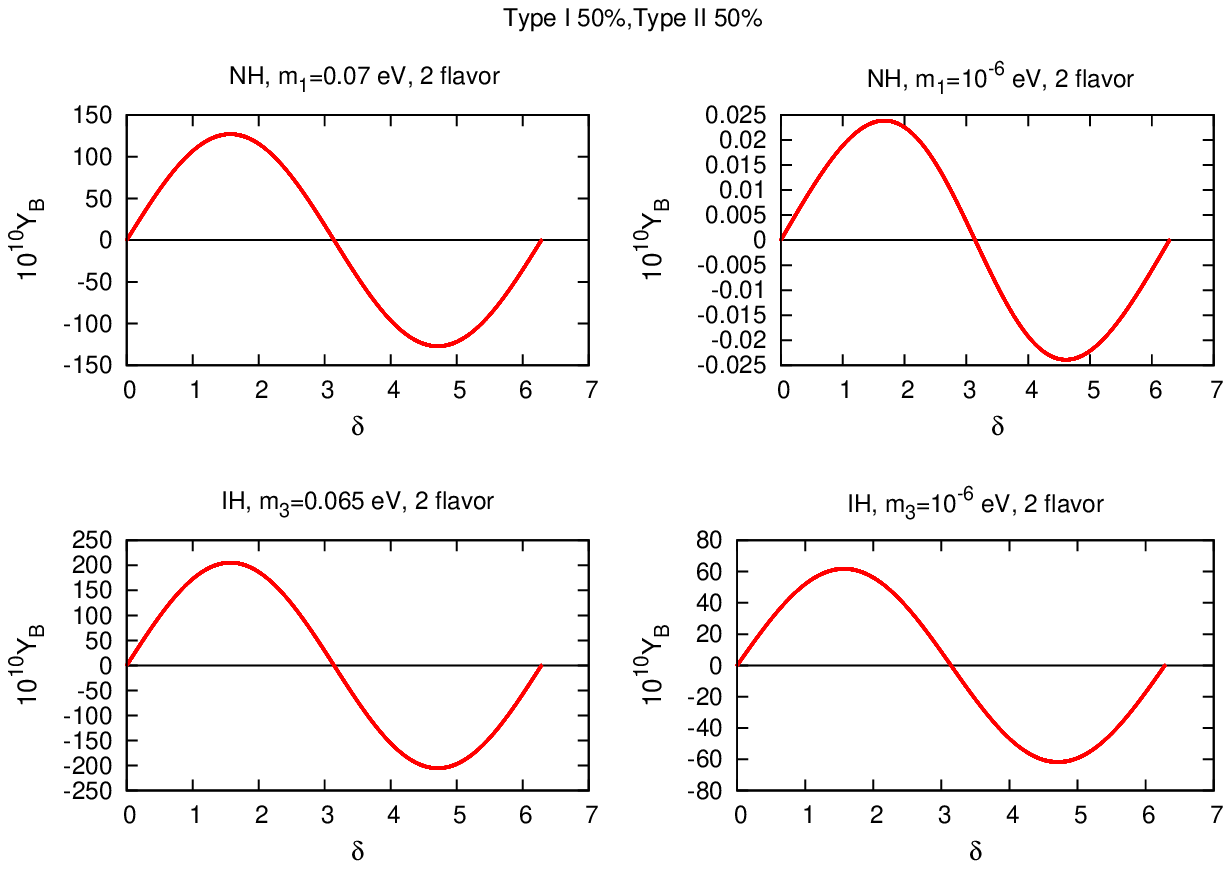}
\caption{Baryon asymmetry as a function of leptonic Dirac CP phase in the two flavor regime for $50\%$ contribution of type I seesaw to neutrino mass}
\label{fig10}
\end{figure}
\begin{figure}[ht]
 \centering
\includegraphics[width=0.85\textwidth]{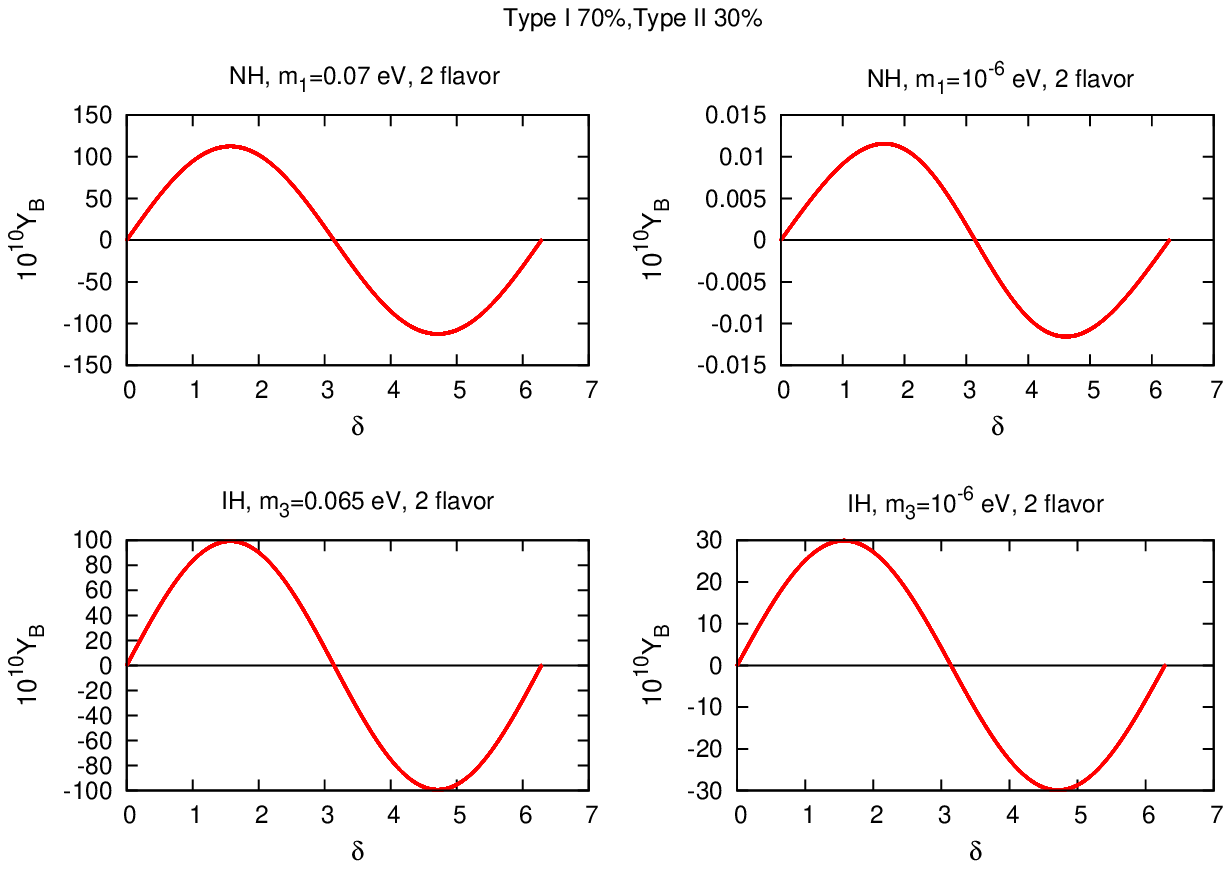}
\caption{Baryon asymmetry as a function of leptonic Dirac CP phase in the two flavor regime for $70\%$ contribution of type I seesaw to neutrino mass}
\label{fig11}
\end{figure}
\begin{figure}[ht]
 \centering
\includegraphics[width=0.85\textwidth]{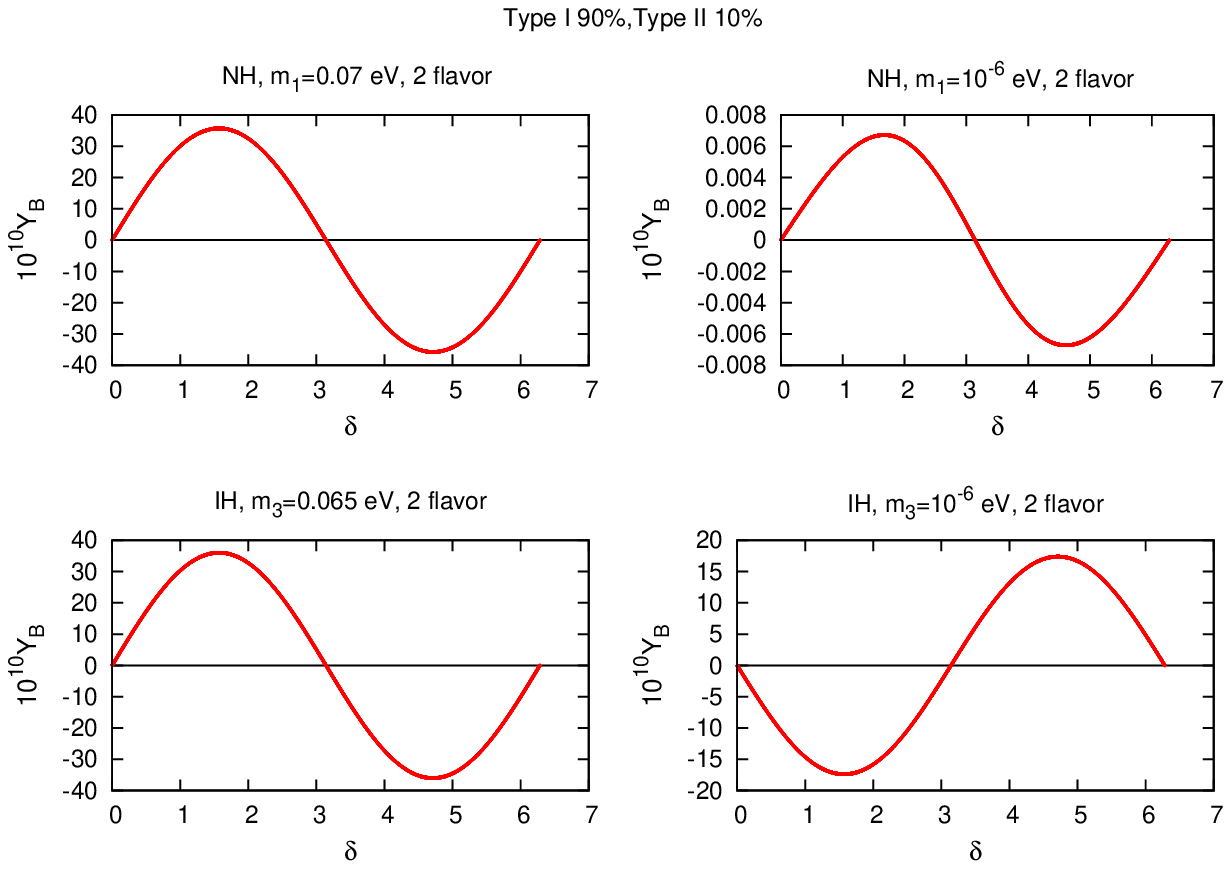}
\caption{Baryon asymmetry as a function of leptonic Dirac CP phase in the two flavor regime for $90\%$ contribution of type I seesaw to neutrino mass}
\label{fig12}
\end{figure}
\begin{figure}[ht]
 \centering
\includegraphics[width=0.85\textwidth]{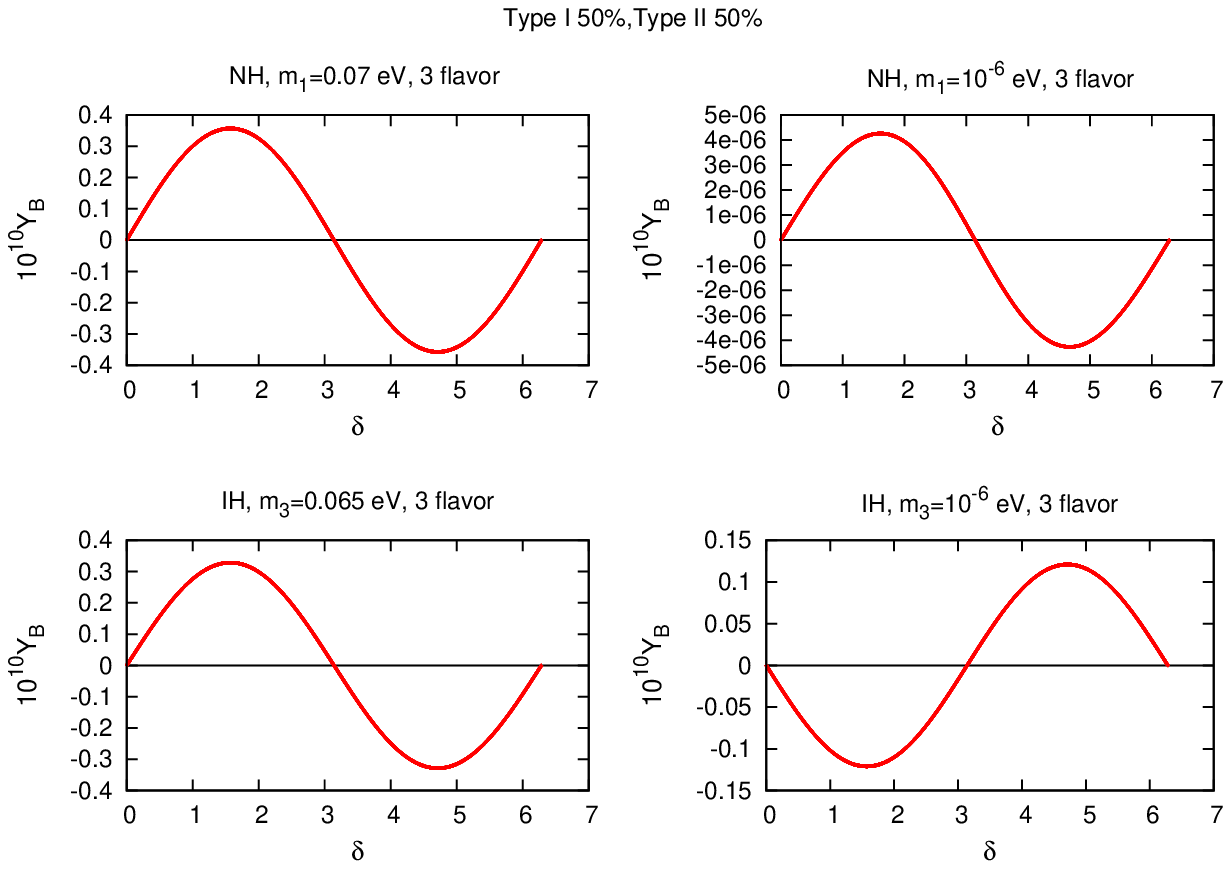}
\caption{Baryon asymmetry as a function of leptonic Dirac CP phase in the three flavor regime for $50\%$ contribution of type I seesaw to neutrino mass}
\label{fig13}
\end{figure}
\begin{figure}[ht]
 \centering
\includegraphics[width=0.85\textwidth]{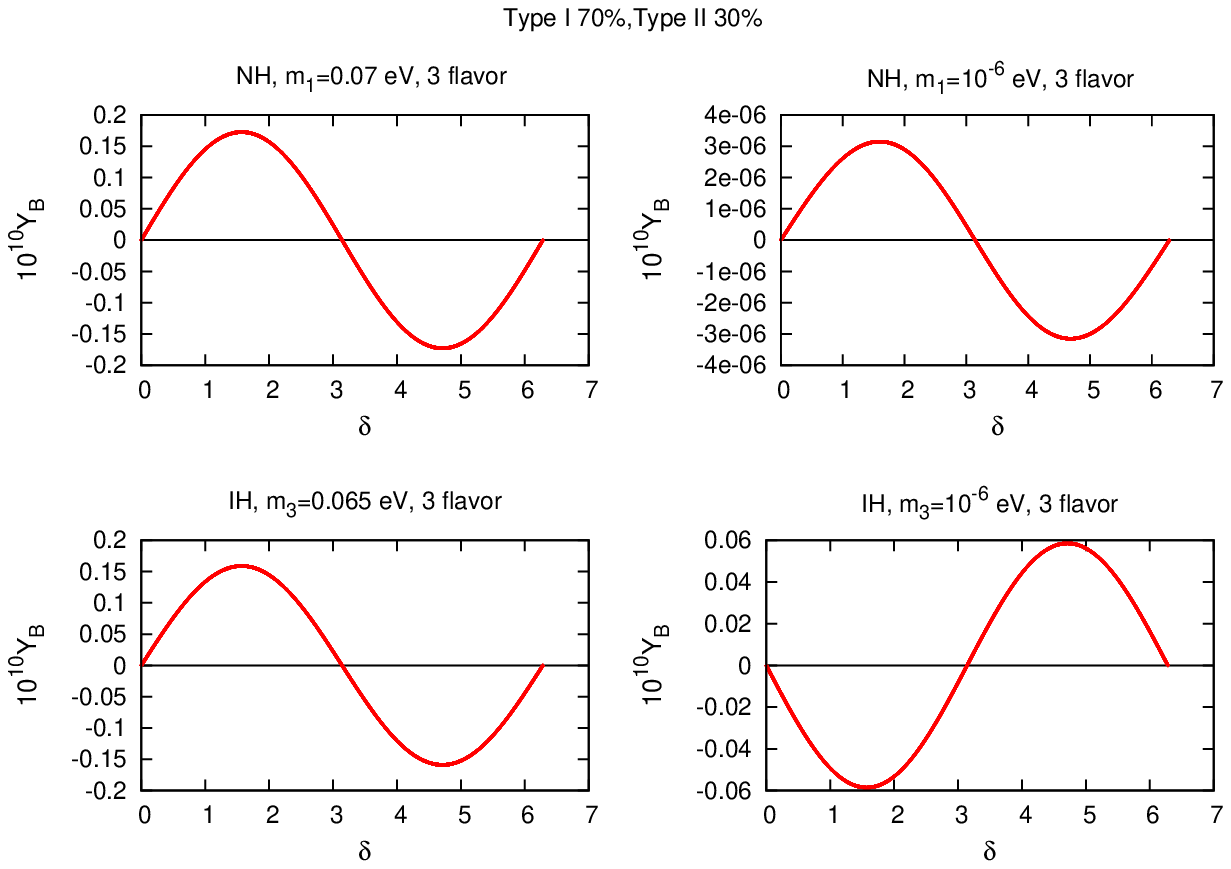}
\caption{Baryon asymmetry as a function of leptonic Dirac CP phase in the three flavor regime for $70\%$ contribution of type I seesaw to neutrino mass}
\label{fig14}
\end{figure}
\begin{figure}[ht]
 \centering
\includegraphics[width=0.85\textwidth]{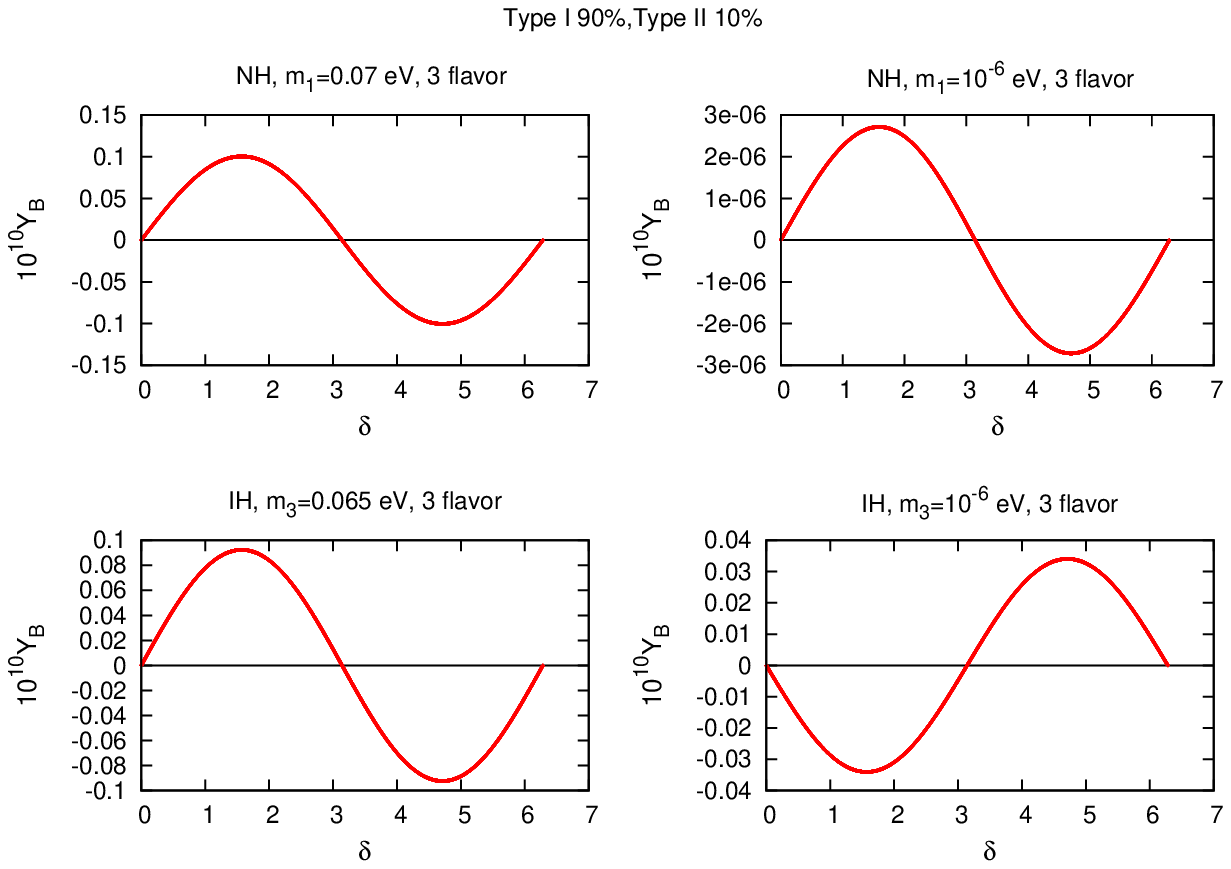}
\caption{Baryon asymmetry as a function of leptonic Dirac CP phase in the three flavor regime for $90\%$ contribution of type I seesaw to neutrino mass}
\label{fig15}
\end{figure}

\section{Deviations from TBM mixing with Type II Seesaw}
\label{sec:devtbm}
As mentioned in the introduction, here we consider the type I seesaw to give rise to TBM type neutrino mixing whereas the type II seesaw term acts like a deviation from TBM mixing in order to generate non-zero $\theta_{13}$. The TBM type neutrino mass matrix can be parametrized as
\begin{equation}
M_{\nu}=\left(\begin{array}{ccc}
x& y&y\\
y& x+z & y-z \\
y & y-z & x+z
\end{array}\right)
\label{matrix1}
\end{equation}
which is clearly $\mu-\tau$ symmetric with eigenvalues $m_1 = x-y, \; m_2 = x+2y, \; m_3 = x-y+2z$. It predicts the mixing angles as $\theta_{12} \simeq 35.3^o, \; \theta_{23} = 45^o$ and $\theta_{13} = 0$. This corresponding mixing matrix can be written as
\begin{equation}
U_{TBM}==\left(\begin{array}{ccc}\sqrt{\frac{2}{3}}&\frac{1}{\sqrt{3}}&0\\
 -\frac{1}{\sqrt{6}}&\frac{1}{\sqrt{3}}&\frac{1}{\sqrt{2}}\\
\frac{1}{\sqrt{6}}&-\frac{1}{\sqrt{3}}& \frac{1}{\sqrt{2}}\end{array}\right)
\end{equation}
The Pontecorvo-Maki-Nakagawa-Sakata (PMNS) leptonic mixing matrix is related to the diagonalizing 
matrices of neutrino and charged lepton mass matrices $U_{\nu}, U_l$ respectively, as
\begin{equation}
U_{\text{PMNS}} = U^{\dagger}_l U_{\nu}
\label{pmns0}
\end{equation}
The PMNS mixing matrix can be parametrized as
\begin{equation}
U_{\text{PMNS}}=\left(\begin{array}{ccc}
c_{12}c_{13}& s_{12}c_{13}& s_{13}e^{-i\delta}\\
-s_{12}c_{23}-c_{12}s_{23}s_{13}e^{i\delta}& c_{12}c_{23}-s_{12}s_{23}s_{13}e^{i\delta} & s_{23}c_{13} \\
s_{12}s_{23}-c_{12}c_{23}s_{13}e^{i\delta} & -c_{12}s_{23}-s_{12}c_{23}s_{13}e^{i\delta}& c_{23}c_{13}
\end{array}\right) 
\label{matrixPMNS}
\end{equation}
where $c_{ij} = \cos{\theta_{ij}}, \; s_{ij} = \sin{\theta_{ij}}$ and $\delta$ is the Dirac CP phase. If $U_{\nu} = U_{TBM}$ from type I seesaw, then for diagonal charged lepton mass matrix both the reactor mixing angle $\theta_{13}$ and the leptonic Dirac CP phase $\delta$ vanish in the neutrino sector. Since vanishing reactor mixing angle has been ruled out by experiments, the TBM type neutrino mixing has to be corrected. In our present work, we are interested in exploring the possibility of generating this correction for non-zero $\theta_{13}$ and predicting leptonic CP phase at the same time. Considering the type II seesaw term as the necessary correction to TBM mixing, we write the neutrino mass matrix as
\begin{equation}
M_{\nu} = M^{II}_{\nu} + M^{I}_{\nu} = f v_{L} + M_{D}M_{R}^{-1}M_{D}^{T}
\end{equation}
where $M_D, M_R$ are Dirac the right handed neutrino mass matrices respectively. Since the diagonalizing matrix of $M_{\nu}$ is $U_{\text{PMNS}}$ and that of type I mass matrix $M^I_{\nu}$ is $U_{TBM}$, the above equation can be written as
\begin{equation}
U_{\text{PMNS}}M^{\text{diag}}_{\nu} U^T_{\text{PMNS}} = M^{II}_{\nu} + U_{TBM} M^{I(\text{diag})}_{\nu} U^T_{TBM}
\label{nu12}
\end{equation}
For normal hierarchy, the diagonal mass matrix of the light neutrinos can be written  as $M^{\text{diag}}_{\nu} 
= \text{diag}(m_1, \sqrt{m^2_1+\Delta m_{21}^2}, \sqrt{m_1^2+\Delta m_{31}^2})$ whereas for inverted hierarchy 
 it can be written as $M^{\text{diag}}_{\nu} = \text{diag}(\sqrt{m_3^2+\Delta m_{23}^2-\Delta m_{21}^2}, 
\sqrt{m_3^2+\Delta m_{23}^2}, m_3)$. To vary the relative strength of type I and type II seesaw terms, we parametrize the diagonal type I mass matrix as $M^{I(\text{diag})}_{\nu} = \alpha M^{\text{diag}}_{\nu}$, where $\alpha$ is a parameter between 0 and 1. Denoting the symmetric type II seesaw mass matrix as
\begin{equation}
M^{II}_{\nu}=\left(\begin{array}{ccc}
t_{11}& t_{12}&t_{13}\\
t_{12}& t_{22} & t_{23} \\
t_{13} & t_{23} & t_{33}
\end{array}\right)
\label{type2matrix}
\end{equation}
and using equation (\ref{nu12}), the type II seesaw mass matrix elements can be derived as
{\small $$t_{11}=\left(c^2_{12}c^2_{13}-\frac{2\alpha}{3}\right)m_1+\left(s^2_{12}c^2_{13}-\frac{\alpha}{3}\right)m_2+s^2_{13}e^{-2i\delta}m_3$$
 $$ t_{12}=\left(-c_{12}c_{13}\left(s_{12}c_{23}+e^{i\delta}s_{13}s_{23}c_{13}\right)+\frac{\alpha}{3}\right)m_1+\left(s_{12}c_{13}\left(c_{12}c_{23}-e^{i\delta}s_{12}s_{13}s_{23}\right)-\frac{\alpha}{3}\right)m_2 +s_{13}s_{23}c_{13}e^{-i\delta}m_3$$

 $$t_{13}=\left(c_{12}c_{13}\left(s_{12}s_{23}-e^{i\delta}s_{13}c_{12}c_{23}\right)+\frac{\alpha}{3}\right)m_1+\left(-s_{12}c_{13}\left(c_{12}s_{23}+e^{i\delta}s_{12}s_{13}c_{23}\right)-\frac{\alpha}{3}\right)m_2\\ +s_{13}c_{13}c_{23}e^{-i\delta}m_3$$

 $$t_{22}=\left( \left(s_{12}c_{23}+e^{i\delta}s_{13}s_{23}c_{12}\right)^{2}-\frac{\alpha}{6}\right)m_1\\+\left(\left(c_{12}c_{23}-e^{i\delta}s_{12}s_{13}s_{23}\right)^2-\frac{\alpha}{3}\right)m_2\\ +\left(s^2_{23}c^2_{13}-\frac{\alpha}{2}\right)m_3$$

\begin{multline*}
 $$ t_{23}=\left(\left(-s_{12}s_{23}+e^{i\delta}s_{13}c_{12}c_{23}\right)\left(s_{12}c_{23}+e^{i\delta}s_{13}s_{23}c_{12}\right)-\frac{\alpha}{6}\right)m_1\\ +\left(-\left(s_{23}c_{12}+e^{i\delta}s_{12}s_{13}c_{23}\right)\left(c_{12}c_{23}-e^{i\delta}s_{12}s_{13}s_{23}\right)-\frac{\alpha}{3}\right)m_2+\left(s_{23}c_{23}c^2_{13}+\frac{\alpha}{2}\right)m_3$$
\end{multline*}

 $$t_{33}=\left(\left(-s_{12}s_{23}+e^{i\delta}s_{13}c_{12}c_{23}\right)^2-\frac{\alpha}{6}\right)m_1\\+\left(\left(s_{23}c_{12}+e^{i\delta}s_{12}s_{13}c_{23}\right)^2-\frac{\alpha}{3}\right)m_2\\ +\left(c^2_{13}c^2_{23}-\frac{\alpha}{2}\right)m_3$$}

It should be noted that in the general PMNS mixing matrix, there are two Majorana phases as well originating in the right handed neutrino sector. We have omitted these two phases in equation \eqref{matrixPMNS}, assuming them to take trivial values $(0, 2\pi)$. These Majorana phases can significantly affect the baryon asymmetry as discussed recently within the context of type I and type II seesaw models in \cite{leptodborah}. The charged lepton flavor violation within such frameworks have been discussed recently in \cite{dbmkdsp} whereas the consequences in neutrinoless double beta decay have been explored in details in \cite{ndbdlr,ndbdlr2}. We leave a more detailed study of these scenarios including the effects of non-trivial Majorana phases to future investigations. 

\section{Numerical Analysis}
\label{sec:numeric}
Using the expressions for type II seesaw mass matrix elements in terms of neutrino parameters, we first try to see the dependence of leptonic Dirac CP phase $\delta = \delta_{CP}$ on type II seesaw strength. For that, we use the best fit values of the neutrino parameters given in \cite{schwetz12}. Since neutrino oscillation data give only two mass squared differences and three mixing angles, we still have the lightest neutrino mass as the free parameter. Thus, the dependence of $\delta_{CP}$ on type II seesaw is subject to the choice of two free parameters: the lightest neutrino mass and the relative strength of type I-II seesaw terms $\alpha$. We choose two different values of lightest neutrino mass which give rise to quasi-degenerate and hierarchical type neutrino mass patterns respectively. We also choose three different values of $\alpha$ to study different relative contribution of type I and type II seesaw terms. The variation of $\sin^2 \delta_{CP}$ on type II seesaw is shown in figure \ref{fig1}, \ref{fig2}, \ref{fig3} for different choices of lightest neutrino mass and $\alpha$. For normal hierarchy, the choices for lightest neutrino mass are $m_1 = 10^{-6}, 0.07$ eV whereas for inverted hierarchy they are taken as $m_3 = 10^{-6}, 0.065$ eV. The specific choice of $m_1 (m_3) = 10^{-6}$ eV corresponds to a large hierarchy in the neutrino mass spectrum and $m_1 (m_3) = 0.07 (0.065)$ eV correspond to a neutrino mass spectrum where all the three masses are of same order. The values $0.07, 0.065$ eV are chosen in such a way that the sum of absolute neutrino masses lies below the cosmological upper bound. We choose three different values of $\alpha$ namely, $0.5, 0.7, 0.9$ which correspond to $50\%, 70\%, 90\%$ contribution of type I seesaw to neutrino mass respectively. We also show the parameter space in the type II seesaw term versus lightest neutrino mass plane which can give rise to the correct neutrino oscillation data within our framework. The corresponding plots for all the relative strengths of type I and type II seesaw terms are shown in figure \ref{fig4}, \ref{fig5} and \ref{fig6}. We also derive the type II seesaw mass matrices for all the choices of lightest neutrino mass and $\alpha$. They are shown in table \ref{table1}, \ref{table2}, \ref{table3} and \ref{table4}.

\begin{table}
\caption{Type II Seesaw Mass matrix for NH with $m_1 = 0.07$ eV}
\begin{center}

\resizebox{\textwidth}{!}{%
\begin{tabular}{ |c| c| c|  }
\hline
\textbf{MODEL(TBM)} & \textbf{NH $m_1=0.07$ eV}\\
\hline

\addlinespace[1.5ex]
\hline
 TYPE I $50\%$ &
$ \begin{pmatrix}0.0328386 + 0.0019316 e^{-2i\delta} &0.0000971076 + 0.00807665 e^{-i \delta}-0.00665413 e^{i \delta} &-0.000243589 + 0.00971721 e^{-i \delta}-0.00800575 e^{i \delta}\\0.0000971076 + 0.00807665 e^{-i \delta}-0.00665413 e^{i \delta} & 0.0367825 - 0.0000364514e^{i \delta}+0.000651833  e^{2i \delta}&0.00954316 - 6.77917\times10^{-6} e^{i \delta} + 0.000784237  e^{2i \delta}\\-0.000243589 + 0.00971721  e^{-i \delta}-0.00800575 e^{i \delta}&0.00954316 - 6.77917\times10^{-6} e^{i \delta} + 0.000784237  e^{2i \delta}&0.0388914 + 0.0000364514 e^{i \delta} + 0.000943535 e^{2i \delta} \end{pmatrix}$\\
\hline
\addlinespace[2.5ex]
\hline
 TYPE I $70\%$ &
$ \begin{pmatrix}0.0188031 + 0.0019316 e^{-2i\delta} &0.000061529 + 0.00807665  e^{-i \delta}-0.00665413 e^{i \delta} &-0.000279168 + 0.00971721e^{-i \delta}-0.00800575 e^{i \delta}\\0.000061529 + 0.00807665  e^{-i \delta}-0.00665413 e^{i \delta} & 0.021162 - 0.0000364514e^{i \delta}+0.000651833  e^{2i \delta}&0.0110924 - 6.77917\times10^{-6} e^{i \delta} + 0.000784237 e^{2i \delta}\\-0.000279168 + 0.00971721e^{-i \delta}-0.00800575 e^{i \delta}&0.0110924 - 6.77917\times10^{-6} e^{i \delta} + 0.000784237 e^{2i \delta}&0.023271 + 0.0000364514 e^{i \delta} + 0.000943535 e^{2i \delta} \end{pmatrix}$\\
\hline
\addlinespace[2.5ex]
\hline
 TYPE I $90\%$ &
$ \begin{pmatrix}0.00476749 + 0.0019316 e^{-2i\delta} &0.0000259503 + 0.00807665  e^{-i \delta}-0.00665413 e^{i \delta} &-0.000314746 + 0.00971721 e^{-i \delta}-0.00800575 e^{i \delta}\\0.0000259503  + 0.00807665 e^{-i \delta}-0.00665413 e^{i \delta} & 0.00554157  - 0.0000364514e^{i \delta}+0.000651833  e^{2i \delta}&0.0126417 - 6.77917\times10^{-6} e^{i \delta} + 0.000784237  e^{2i \delta}\\-0.000314746 + 0.00971721 e^{-i \delta}-0.00800575 e^{i \delta}&0.0126417 - 6.77917\times10^{-6} e^{i \delta} + 0.000784237  e^{2i \delta}&0.00765053 + + 0.0000364514 e^{i \delta} + 0.000943535 e^{2i \delta} \end{pmatrix}$\\
\hline
\end{tabular}
}
\end{center}
\label{table1}
\end{table}

\begin{table}
\caption{Type II Seesaw Mass matrix for NH with $m_1 = 10^{-6}$ eV}
\begin{center}

\resizebox{\textwidth}{!}{%
\begin{tabular}{ |c| c| c|  }
\hline
\textbf{MODEL(TBM)} & \textbf{NH $m_1=10^{-6}$ eV}\\
\hline
\addlinespace[1.5ex]
\hline
 TYPE I $50\%$ &
$ \begin{pmatrix}0.00107295 + 0.00111823 e^{-2i\delta} &0.00157563 + 0.00467569 e^{-i \delta}-0.00024653 e^{i \delta} &-0.00395237 + 0.00562544  e^{-i \delta}-0.000296607  e^{i \delta}\\0.00157563 + 0.00467569 e^{-i \delta}-0.00024653 e^{i \delta} & 0.00930559 - 0.000591445e^{i \delta}+0.0000241499  e^{2i \delta}&0.0314917 - 0.000109996 e^{i \delta} + 0.0000290554   e^{2i \delta}\\-0.00395237 + 0.00562544  e^{-i \delta}-0.000296607  e^{i \delta}&0.0314917 - 0.000109996 e^{i \delta} + 0.0000290554   e^{2i \delta}&0.0169345 + 0.000591445 e^{i \delta} + 0.0000349572 e^{2i \delta} \end{pmatrix}$\\
\hline
\addlinespace[2.5ex]
\hline
 TYPE I $70\%$ &
$ \begin{pmatrix}0.000495471 + 0.00111823 e^{-2i\delta} &0.000998342 + 0.00467569  e^{-i \delta}-0.00024653 e^{i \delta} &-0.00452965 + 0.00562544  e^{-i \delta}-0.000296607  e^{i \delta}\\0.000998342 + 0.00467569  e^{-i \delta}-0.00024653 e^{i \delta} & 0.00375829 - 0.000591445e^{i \delta}+0.0000241499  e^{2i \delta}&0.0358842 - 0.000109996  e^{i \delta} + 0.0000290554   e^{2i \delta}\\-0.00452965 + 0.00562544  e^{-i \delta}-0.000296607  e^{i \delta}&0.0358842 - 0.000109996  e^{i \delta} + 0.0000290554   e^{2i \delta}&0.0113872 + 0.000591445 e^{i \delta} + 0.0000349572 e^{2i \delta} \end{pmatrix}$\\
\hline
\addlinespace[2.5ex]
\hline
 TYPE I $90\%$ &
$ \begin{pmatrix}-0.0000820129 + 0.00111823  e^{-2i\delta} &0.000421058 + 0.00467569 e^{-i \delta}-0.00024653 e^{i \delta} &-0.00510694 + 0.00562544   e^{-i \delta}-0.000296607  e^{i \delta}\\0.000421058 + 0.00467569 e^{-i \delta}-0.00024653 e^{i \delta} & -0.001789 - 0.000591445e^{i \delta}+0.0000241499  e^{2i \delta}&0.0402767 - 0.000109996  e^{i \delta} + 0.0000290554   e^{2i \delta}\\-0.00510694 + 0.00562544   e^{-i \delta}-0.000296607  e^{i \delta}&0.0402767 - 0.000109996  e^{i \delta} + 0.0000290554   e^{2i \delta}&0.00583995 + 0.000591445 e^{i \delta} + 0.0000349572 e^{2i \delta} \end{pmatrix}$\\
\hline
\end{tabular}
}
\end{center}
\label{table2}
\end{table}

\begin{table}
\caption{Type II Seesaw Mass matrix for IH with $m_3=0.065$ eV}
\begin{center}

\resizebox{\textwidth}{!}{%
\begin{tabular}{ |c |c| c|  }
\hline
\textbf{MODEL(TBM)} & \textbf{IH $m_3=0.065$ eV}\\
\hline
\addlinespace[1.5ex]
\hline
 TYPE I $50\%$ &
$ \begin{pmatrix}0.0380633 + 0.0014625 e^{-2i\delta} &0.0000838167 + 0.0061152 e^{-i \delta}-0.00771227  e^{i \delta} &-0.00021025 + 0.00735735  e^{-i \delta}-0.00927883 e^{i \delta}\\0.0000838167 + 0.0061152 e^{-i \delta}-0.00771227  e^{i \delta} &0.0376617 - 0.0000314624e^{i \delta}+0.000755488 e^{2i \delta}&-0.0138471 - 5.85132\times10^{-6}  e^{i \delta} +0.000908947 e^{2i \delta}\\-0.00021025 + 0.00735735  e^{-i \delta}-0.00927883 e^{i \delta}&-0.0138471 - 5.85132\times10^{-6}  e^{i \delta} +0.000908947 e^{2i \delta}&0.0340442 + 0.0000314624 e^{i \delta} + 0.00109358 e^{2i \delta} \end{pmatrix}$\\
\hline
\addlinespace[2.5ex]
\hline
 TYPE I $70\%$ &
$ \begin{pmatrix}0.0217968 + 0.0014625 e^{-2i\delta} &0.0000531076 + 0.0061152 e^{-i \delta}-0.00771227  e^{i \delta} &-0.000240959 + 0.00735735 e^{-i \delta}-0.00927883 e^{i \delta}\\0.0000531076 + 0.0061152 e^{-i \delta}-0.00771227  e^{i \delta} &0.0230131 - 0.0000314624 e^{i \delta}+0.000755488 e^{2i \delta}&-0.0154957 - 5.85132\times10^{-6}  e^{i \delta} +0.000908947 e^{2i \delta}\\-0.000240959 + 0.00735735 e^{-i \delta}-0.00927883 e^{i \delta}&-0.0154957 - 5.85132\times10^{-6}  e^{i \delta} +0.000908947 e^{2i \delta}&0.0193956 + 0.0000314624 e^{i \delta} + 0.00109358 e^{2i \delta} \end{pmatrix}$\\
\hline
\addlinespace[2.5ex]
\hline
 TYPE I $90\%$ &
$ \begin{pmatrix}0.00553034 + 0.0014625  e^{-2i\delta} &0.0000223985 + 0.0061152 e^{-i \delta}-0.00771227  e^{i \delta} &-0.000271668 + 0.00735735  e^{-i \delta}-0.00927883 e^{i \delta}\\0.0000223985 + 0.0061152 e^{-i \delta}-0.00771227  e^{i \delta} &0.00836452 - 0.0000314624e^{i \delta}+0.000755488 e^{2i \delta}&-0.0171443 - 5.85132\times10^{-6}  e^{i \delta} +0.000908947 e^{2i \delta}\\-0.000271668 + 0.00735735  e^{-i \delta}-0.00927883 e^{i \delta}&-0.0171443 - 5.85132\times10^{-6}  e^{i \delta} +0.000908947 e^{2i \delta}&0.00474702 + 0.0000314624 e^{i \delta} + 0.00109358 e^{2i \delta} \end{pmatrix}$\\
\hline
\end{tabular}
}
\end{center}
\label{table3}
\end{table}

\begin{table}
\caption{Type II Seesaw Mass Matrix for IH with $m_3=10^{-6}$ eV}
\begin{center}

\resizebox{\textwidth}{!}{%
\begin{tabular}{ |c| c| c|  }
\hline
\textbf{MODEL(TBM)} & \textbf{IH $m_3=10^{-6}$ eV}\\
\hline
\addlinespace[1.5ex]
\hline
 TYPE I $50\%$ &
$ \begin{pmatrix}0.0228629 + 2.25\times10^{-8} e^{-2i\delta} &0.000139215 + 9.408\times10^{-8} e^{-i \delta}-0.00463406   e^{i \delta} &-0.000349213 + 1.1319\times10^{-7} e^{-i \delta}-0.00557536e^{i \delta}\\0.000139215 + 9.408\times10^{-8} e^{-i \delta}-0.00463406   e^{i \delta} &0.0171019 - 0.0000522573e^{i \delta}+0.000453949 e^{2i \delta}&-0.0367103 - 9.71871\times10^{-6}  e^{i \delta} +0.000546157 e^{2i \delta}\\-0.000349213 + 1.1319\times10^{-7} e^{-i \delta}-0.00557536e^{i \delta}&-0.0367103 - 9.71871\times10^{-6}  e^{i \delta} +0.000546157 e^{2i \delta}&0.00801682 + 0.0000522573 e^{i \delta} + 0.000657096  e^{2i \delta} \end{pmatrix}$\\
\hline
\addlinespace[2.5ex]
\hline
 TYPE I $70\%$ &
$ \begin{pmatrix}0.0130856  + 2.25\times10^{-8} e^{-2i\delta} &0.0000882087  + 9.408\times10^{-8} e^{-i \delta}-0.00463406   e^{i \delta} &-0.000400219 + 1.1319\times10^{-7} e^{-i \delta}-0.00557536e^{i \delta}\\0.0000882087 + 9.408\times10^{-8} e^{-i \delta}-0.00463406   e^{i \delta} &0.0121877  - 0.0000522573e^{i \delta}+0.000453949 e^{2i \delta}&-0.0416243 - 9.71871\times10^{-6}  e^{i \delta} +0.000546157 e^{2i \delta}\\-0.000400219 + 1.1319\times10^{-7} e^{-i \delta}-0.00557536e^{i \delta}&-0.0416243 - 9.71871\times10^{-6}  e^{i \delta} +0.000546157 e^{2i \delta}&0.00310259  + 0.0000522573 e^{i \delta} + 0.000657096  e^{2i \delta} \end{pmatrix}$\\
\hline
\addlinespace[2.5ex]
\hline
 TYPE I $90\%$ &
$ \begin{pmatrix}0.00330839 + 2.25\times10^{-8} e^{-2i\delta} &0.0000372027  + 9.408\times10^{-8} e^{-i \delta}-0.00463406   e^{i \delta} &-0.000451225 + 1.1319\times10^{-7} e^{-i \delta}-0.00557536e^{i \delta}\\0.0000372027 + 9.408\times10^{-8} e^{-i \delta}-0.00463406   e^{i \delta} &0.00727345 - 0.0000522573e^{i \delta}+0.000453949 e^{2i \delta}&-0.0465383 - 9.71871\times10^{-6}  e^{i \delta} +0.000546157 e^{2i \delta}\\-0.000451225 + 1.1319\times10^{-7} e^{-i \delta}-0.00557536e^{i \delta}&-0.0465383 - 9.71871\times10^{-6}  e^{i \delta} +0.000546157 e^{2i \delta}&-0.00181164 + 0.0000522573 e^{i \delta} + 0.000657096  e^{2i \delta} \end{pmatrix}$\\
\hline
\end{tabular}
}
\end{center}
\label{table4}
\end{table}

After deriving the type II seesaw mass matrix, we then proceed to calculate the amount of baryon asymmetry that can be produced through the mechanism of leptogenesis in our model. Since type I seesaw term gives rise to a real TBM type neutrino mass matrix, it is natural to assume real values of Dirac Yukawa couplings in the terms $Y_{ij} \bar{L_i} H N_j$. This makes the process of right handed heavy neutrino decay into a light neutrino and Higgs $(N \rightarrow \nu H)$ CP preserving ruling out the possibility of leptogenesis. However, the type II seesaw mass matrix, being the common origin of non-zero $\theta_{13}$ and leptonic Dirac CP phase, is of complex type and hence could be a source of lepton asymmetry. Type II seesaw or the presence of Higgs triplet can contribute to lepton asymmetry in two different ways: CP violating decay of triplet Higgs into standard model particles or the right handed neutrino decay into standard model particles through virtual Higgs triplet. We adopt the latter possibility where CP violating decay of the lightest right handed neutrino through a virtual Higgs triplet is responsible for creating the lepton asymmetry. To calculate the baryon asymmetry, we follow the procedure adopted in our earlier works \cite{dbijmpa, dbmkdsp, leptodborah}. 

While calculating the lepton asymmetry, the lepton flavor effects \cite{flavorlepto,flavorlepto2,flavorlepto3,flavorlepto4,flavorlepto5} can play a crucial role if the lightest right handed neutrino mass $M_1$ is below $10^{12}$ GeV. For $M_1 > 10^{12}$ GeV, all lepton flavors are out of equilibrium giving rise to one flavor leptogenesis. If $10^9 \text{GeV} < M_1 < 10^{12} \text{GeV}$, then only electron and muon flavors are out of equilibrium giving rise to two flavor leptogenesis. Three flavor leptogenesis arises when $M_1 < 10^9$ GeV where only electron flavor is out of equilibrium. To keep the lightest right handed neutrino mass in different flavor regime, one needs to choose the Dirac neutrino mass matrix appropriately. After fitting the type I seesaw mass matrix as $M^I_{\nu} = U_{TBM} M^{I(\text{diag})}_{\nu} U^T_{TBM}$ where $M^{I(\text{diag})}_{\nu} = \alpha M^{(\text{diag})}_{\nu}$ we can find the right handed neutrino mass matrix as
\begin{equation}
M_{R}=M_{D}^T (M^I_{\nu})^{-1}M_{D}
\label{inverseI}
\end{equation}
We consider a diagonal type Dirac neutrino mass matrix which can be parametrized as
\begin{equation}
M_{D}=\left(\begin{array}{ccc}
\lambda^m & 0 & 0\\
0 & \lambda^n & 0 \\
0 & 0 & 1
\end{array}\right)m_f
\label{mLR1}
\end{equation}
where $\lambda = 0.22$ is the standard Wolfenstein parameter and $(m,n)$ are positive integers. We fix $m_f = 82.43$ GeV and change the integers $(m,n)$ in order to keep the lightest right handed neutrino mass $M_1$ in the appropriate flavor regime. We find that $(m,n)= (1,1), (3,1), (5,3)$ correspond to one, two and three flavor regimes of leptogenesis respectively. After fixing $\alpha, m_1 (m_3), M_{D}, M_R$ the only free parameter in the neutrino sector within our framework is the leptonic CP phase $\delta$. We vary the CP phase and compute the predictions for baryon asymmetry in one, two and three flavor regimes for all the choices of $\alpha, m_1 (m_3)$. The variations of baryon asymmetry with leptonic CP phase $\delta$ are shown in figure \ref{fig7}, \ref{fig8}, \ref{fig9}, \ref{fig10}, \ref{fig11}, \ref{fig12}, \ref{fig13}, \ref{fig14} and \ref{fig15}. 

\begin{table}[h]
\caption{Values of $\delta$ giving correct baryon asymmetry}
\begin{center}

\begin{tabular}{|l|l|l|l|}
\hline
\multicolumn{2}{|l|}{Model (TBM)} & $\delta $ for 1 flavor (radian) &$\delta $ for 2 flavor (radian)  \\ \hline
\hline
\multirow{3}{*}{TYPE I 50\%}  & $m_1=0.070$ eV (NH)  & 0.0025-0.0031,3.138-3.139 &0.0433-0.0546,3.089-3.097  \\ \cline{2-4} 
                   & $m_1=10^{-6}$ eV (NH)  & - &- \\ \cline{2-4} 
                  & $m_3=0.065$ eV (IH)  & - & 0.0270-0.0339,3.108-3.114 \\ \cline{2-4} 
                   & $m_3=10^{-6}$ eV (IH)   &-  &0.0898-0.1124,3.035-3.052 \\ \hline
                   \hline
\multirow{3}{*}{TYPE I 70\%}  & $m_1=0.070$ eV (NH)   &3.1470-3.1490,6.276-6.277 & 0.0490-0.0615,3.082-3.092 \\ \cline{2-4} 
& $m_1=10^{-6}$ eV (NH)  & - &- \\ \cline{2-4} 
                   &   $m_3=0.065$ eV (IH)  &-  & 0.0550-0.0700,3.075-3.085 \\ \cline{2-4} 
                   & $m_3=10^{-6}$ eV (IH)  & - &0.1853-0.2343,2.919-2.954  \\ \hline
                   \hline
\multirow{3}{*}{TYPE I 90\%}  & $m_1=0.070$ eV (NH)  &0.0087-0.0119,3.129-3.132  & 0.1545-0.1960,2.955-2.983 \\ \cline{2-4} 
& $m_1=10^{-6}$ eV (NH)  & - &- \\ \cline{2-4} 
                   & $m_3=0.065$ eV (IH)  & - &0.1545-0.1954,2.958-2.985  \\ \cline{2-4} 
                   & $m_3=10^{-6}$ eV (IH)  & - &3.4670-3.5540,5.898-5.960  \\ \hline
                
\end{tabular}
   \end{center}
\label{table5}
\end{table}
\section{Results and Conclusion}
\label{sec:conclude}
In this work, we have generalized our earlier studies on the common origin of non-zero reactor mixing angle, leptonic CP phase and leptogenesis from type II seesaw. Type I seesaw is assumed to give a TBM type neutrino mixing whereas type II seesaw gives the necessary corrections to generate non-zero $\theta_{13}$ as well as leptonic CP phase $\delta$. Without assuming any specific form of the type II seesaw mass matrix, here we derive the most general form of type II seesaw mass matrix that can generate non-zero values of $\theta_{13}, \delta$. Unlike previous studies, here we do not assume the type I seesaw as the leading contributor to neutrino mass. We consider three different scenarios where type I seesaw contributions to neutrino mass are $50\%, 70\%$ and $90\%$ respectively. We also consider both inverted and normal hierarchical neutrino mass spectrum as well as two different types of lightest neutrino mass in order to show the effect of hierarchy. We first see the dependence of leptonic CP phase on type II seesaw mass term. From the figures \ref{fig1}, \ref{fig2}, \ref{fig3} we see that $\sin^2\delta$ varies between 0 and 1 depending on the strength of the type II seesaw term. The variation is found to be very sharp in case of inverted hierarchy with $m_3=10^{-6}$ eV whereas in other cases it varies moderately. We also plot the allowed values of type II seesaw term consistent with neutrino oscillation data as a function of lightest neutrino mass. They are shown in figures \ref{fig4}, \ref{fig5} and \ref{fig6}. We show the variation of baryon asymmetry with leptonic CP phase in figures \ref{fig7}, \ref{fig8}, \ref{fig9}, \ref{fig10}, \ref{fig11}, \ref{fig12}, \ref{fig13}, \ref{fig14} and \ref{fig15} and summarize them in the table \ref{table5}. It can be seen from these figures that we do not get the correct baryon asymmetry in the three flavor regime for all choices of parameters. In the one flavor regime, only normal hierarchical neutrino mass with all masses of same order gives the correct baryon asymmetry. In the two flavor regime although normal hierarchy gives correct baryon asymmetry only when all the neutrino masses are of same order, inverted hierarchy can give rise to correct baryon asymmetry even if the lightest neutrino mass is as low as $10^{-6}$ eV. Precise determination of leptonic CP violation in future experiments should be able to rule out some of the scenarios we have discussed in this work. If we go by the best fit values of $\delta$ presently available in the literature: $5\pi/3$ \cite{schwetz12} and $\pi$ \cite{fogli}, we see from table \ref{table5} that only normal hierarchy with $m_1 = 0.07$ eV prefers the value of $\delta$ to be close to $\pi$ in order to produce the correct baryon asymmetry. On the other hand, none of the scenarios we discuss here favor the value of $\delta$ to be $5\pi/3$. Apart from neutrino experiments, the present and future cosmology experiments should also be able to rule out some more regions of the parameter space we have studied in this work. As stressed earlier, our main conclusions in this work are valid only when the Majorana neutrino phases take only extremal values like $0$ or $2\pi$. We leave a more complete study of these scenarios including the Majorana phases to future works.

\end{document}